\documentclass[ALICE,manyauthors]{cernphprep}

\RequirePackage{ifpdf}
\usepackage{graphicx}
\usepackage{amsmath}
\usepackage{amssymb}
\usepackage{xspace}
\usepackage{float}
\usepackage{placeins}
\usepackage{lineno}
\usepackage[pdftex,usenames,dvipsnames]{xcolor}
\usepackage{cite}
\usepackage{pdflscape}
\usepackage{ctable}
\usepackage{rotating}
\usepackage{multirow}
\usepackage{tabulary}
\usepackage{array,booktabs}
\usepackage{tabularx}
\newcolumntype{Y}{>{\centering\arraybackslash}X}
\ifpdf
  \DeclareGraphicsExtensions{.pdf}
  \usepackage[pdftex]{hyperref}

  \hypersetup{
    final,
    a4paper=true,
    colorlinks=false,
    bookmarks=true,
    pdffitwindow=true,
    pdfnewwindow=true,	
    pdfauthor={ALICE Collaboration},
    pdfcreator={pdflatex},
    pdfproducer={ALICE Collaboration},
    pdftitle={Title},
    pdfsubject={Title},
    pdfkeywords={}
  }

\else
  \usepackage[pdftex]{hyperref}
\fi
\usepackage{hyperref}

\usepackage{cleveref}

\crefname{table}{Tab.}{Tabs.}

\begin{document}
\renewcommand{\figurename}{Figure}
\newcommand {\red}[1]    {\textcolor{red}{#1}}
\newcommand {\green}[1]  {\textcolor{green}{#1}}
\newcommand {\blue}[1]   {\textcolor{blue}{#1}}
\newcommand {\dgreen}[1]    {\textcolor{darkgreen}{#1}}
\newcommand {\orange}[1]   {\textcolor{orange}{#1}}

\newcommand{\deutsch}[1]{\foreignlanguage{german}{#1}}

\newcommand{\CZ}[1]{\blue{{\textbf{CZ:} #1}}}
\newcommand{\PA}[1]{\red{{\textbf{PA:} #1}}}
\newcommand{\JN}[1]{\red{{\textbf{JN:} #1}}}
\newcommand{\MF}[1]{\red{{\textbf{JN:} #1}}}
\newcommand{\ToDo}      {\textcolor{blue}{\footnotesize \textsc{ToDo}}}
\newcommand{\TBC}       {\textcolor{red}{\footnotesize \textsc{TBC}}}

\DeclareRobustCommand{\unit}[2][]{%
        \begingroup%
                \def\0{#1}%
                \expandafter%
        \endgroup%
        \ifx\0\@empty%
                \ensuremath{\mathrm{#2}}%
        \else%
                \ensuremath{#1\,\mathrm{#2}}%
        \fi%
        }
\DeclareRobustCommand{\unitfrac}[3][]{%
        \begingroup%
                \def\0{#1}%
                \expandafter%
        \endgroup%
        \ifx\0\@empty%
                \raisebox{0.98ex}{\ensuremath{\mathrm{\scriptstyle#2}}}%
                \nobreak\hspace{-0.15em}\ensuremath{/}\nobreak\hspace{-0.12em}%
                \raisebox{-0.58ex}{\ensuremath{\mathrm{\scriptstyle#3}}}%
        \else
                \ensuremath{#1}\,%
                \raisebox{0.98ex}{\ensuremath{\mathrm{\scriptstyle#2}}}%
                \nobreak\hspace{-0.15em}\ensuremath{/}\nobreak\hspace{-0.12em}%
                \raisebox{-0.58ex}{\ensuremath{\mathrm{\scriptstyle#3}}}%
        \fi%
}

%
%
\newcommand{\ie}{i.\,e.\;}
\newcommand{\eg}{e.\,g.\;}

\newcommand{\run}[1]{\textsc{Run\,#1}}

%
%

\newcommand{\dNdeta}{\ensuremath{\mathrm{d}N_{\rm ch}/\mathrm{d}\eta}\xspace}

%
%
\newlength{\smallerpicsize}
\setlength{\smallerpicsize}{70mm}
\newlength{\smallpicsize}
\setlength{\smallpicsize}{90mm}
\newlength{\mediumpicsize}
\setlength{\mediumpicsize}{120mm}
\newlength{\largepicsize}
\setlength{\largepicsize}{150mm}

\newcommand{\PICX}[5]{
   \begin{figure}[!hbt]
      \begin{center}
         \vspace{3ex}
         \includegraphics[width=#3]{#1}
         \caption[#4]{\label{#2} #5}        
      \end{center}  
   \end{figure}
}

\newcommand{\PICH}[5]{
   \begin{figure}[H]
      \begin{center}
         \vspace{3ex}
         \includegraphics[width=#3]{#1}
         \caption[#4]{\label{#2} #5}        
      \end{center}  
   \end{figure}
}

%
%
%
\newcommand{\figs}{Figs.\xspace}
\newcommand{\Figs}{Figures\xspace}
\newcommand{\eqn}{equation\xspace}
\newcommand{\Eqn}{Equation\xspace}
\newcommand{\figref}[1]{Fig.~\ref{#1}}
\newcommand{\Figref}[1]{Figure~\ref{#1}}
\newcommand{\tabref}[1]{Tab.~\ref{#1}}
\newcommand{\Tabref}[1]{Table~\ref{#1}}
\newcommand{\appref}[1]{appendix~\ref{#1}}
\newcommand{\Appref}[1]{Appendix~\ref{#1}}
\newcommand{\secs}{Secs.\xspace}
\newcommand{\Secs}{Sections\xspace}
\newcommand{\secref}[1]{Sec.~\ref{#1}}
\newcommand{\Secref}[1]{Section~\ref{#1}}
\newcommand{\chaps}{Chaps.\xspace}
\newcommand{\Chaps}{Chapters\xspace}
\newcommand{\chapref}[1]{Chap.~\ref{#1}}
\newcommand{\Chapref}[1]{Chapter~\ref{#1}}
\newcommand{\lstref}[1]{Listing~\ref{#1}}
\newcommand{\Lstref}[1]{Listing~\ref{#1}}
%
%
\newcommand{\otoprule}{\midrule[\heavyrulewidth]}
\topfigrule
%
\newcommand {\stat}     {({\it stat.})~}
\newcommand {\syst}     {({\it syst.})~}
 \newcommand {\mom}       {\ensuremath{p}}
\newcommand {\pT}        {\pt}
\newcommand {\meanpT}    {\ensuremath{\langle p_{\mathrm{T}} \kern-0.1em\rangle}\xspace}
\newcommand {\mean}[1]   {\ensuremath{\langle #1 \kern-0.1em\rangle}\xspace} 
\newcommand {\sqrtsNN}   {\ensuremath{\sqrt{s_{\textsc{NN}}}}\xspace}
\newcommand {\sqrts}     {\ensuremath{\sqrt{s}}\xspace}
\newcommand {\vf}        {\ensuremath{v_{\mathrm{2}}}\xspace}
\newcommand {\et}        {\ensuremath{E_{\mathrm{t}}}\xspace}
\newcommand {\mT}        {\ensuremath{m_{\mathrm{T}}}\xspace}
\newcommand {\mTmZero}   {\ensuremath{m_{\mathrm{T}} - m_0}\xspace}
\newcommand {\minv}      {\mbox{$m_{\ee}$}}
\newcommand {\rap}       {\mbox{$y$}}
\newcommand {\absrap}    {\mbox{$\left | y \right | $}}
\newcommand {\rapXi}     {\mbox{$\left | y(\rmXi) \right | $}}
\newcommand {\abspseudorap} {\mbox{$\left | \eta \right | $}}
\newcommand {\pseudorap} {\mbox{$\eta$}}
\newcommand {\cTau}      {\ensuremath{c\tau}}
\newcommand {\sigee}     {$\sigma_E$/$E$}
\newcommand {\dNdy}      {\ensuremath{\mathrm{d}N/\mathrm{d}y}}
\newcommand {\dNdpt}     {\ensuremath{\mathrm{d}N/\mathrm{d}\pT }}
\newcommand {\dNdptdy}   {\ensuremath{\mathrm{d^{2}}N/\mathrm{d}\pT\mathrm{d}y }}
\newcommand {\fracdNdptdy}   {\ensuremath{ \frac{\mathrm{d^{2}}N}{\mathrm{d}\pT\mathrm{d}y } }}
\newcommand {\dNdmtdy}   {\ensuremath{\mathrm{d^{2}}N/\mathrm{d}\mT\mathrm{d}y }}
\newcommand {\dN}        {\ensuremath{\mathrm{d}N }}
\newcommand {\dNsquared} {\ensuremath{\mathrm{d^{2}}N }}
\newcommand {\dpt}       {\ensuremath{\mathrm{d}\pT }}
\newcommand {\dy}        {\ensuremath{\mathrm{d}y}}
\newcommand {\dNdyBold}  {\ensuremath{\boldsymbol{\dN/\dy}}\xspace}
\newcommand {\dNchdy}    {\ensuremath{\mathrm{d}N_\mathrm{ch}/\mathrm{d}y }\xspace}
\newcommand {\dNchdeta}  {\ensuremath{\mathrm{d}N_\mathrm{ch}/\mathrm{d}\eta }\xspace}
\newcommand {\dNchdptdeta}  {\ensuremath{\mathrm{d}N_\mathrm{ch}/\mathrm{d}\pT\mathrm{d}\eta }\xspace}
\newcommand {\Raa}       {\ensuremath{R_\mathrm{AA}}}
\newcommand {\RpPb}       {\ensuremath{R_\mathrm{pPb}}\xspace}
\newcommand {\Nevt}      {\ensuremath{N_\mathrm{evt}}}
\newcommand {\NevtINEL}  {\ensuremath{N_\mathrm{evt}(\textsc{inel})}}
\newcommand {\NevtNSD}   {\ensuremath{N_\mathrm{evt}(\textsc{nsd})}}
\newcommand{\dEdx}       {\ensuremath{\mathrm{d}E/\mathrm{d}x}\xspace}
\newcommand{\ttof}       {\ensuremath{t_\mathrm{TOF}}\xspace}
\newcommand {\ee}        {\mbox{$\mathrm {e^+e^-}$}\xspace}
\newcommand {\ep}        {\mbox{$\mathrm {e\kern-0.05em p}$}\xspace}
\newcommand {\pp}        {\mbox{$\mathrm {p\kern-0.05em p}$}\xspace}
\newcommand {\ppBoldMath} {\mbox{$\mathrm { \mathbf p\kern-0.05em \mathbf p }$}\xspace}
\newcommand {\ppbar}     {\mbox{$\mathrm {p\overline{p}}$}\xspace}
\newcommand {\PbPb}      {\ensuremath{\mbox{Pb--Pb}}\xspace}
\newcommand {\AuAu}      {\ensuremath{\mbox{Au--Au}}\xspace}
\newcommand {\CuCu}      {\ensuremath{\mbox{Cu--Cu}}\xspace}
\renewcommand {\AA}      {\ensuremath{\mbox{A--A}}\xspace}
\newcommand {\pA}        {\ensuremath{\mbox{p--A}}\xspace}
\newcommand {\pPb}       {\ensuremath{\mbox{p--Pb}}\xspace}
\newcommand {\Pbp}       {\ensuremath{\mbox{Pb--p}}\xspace}
\newcommand {\hPM}       {\ensuremath{h^{\pm}}\xspace}
\newcommand {\rphi}      {\ensuremath{(r,\phi)}\xspace}
\newcommand {\alphaS}    {\ensuremath{ \alpha_s}\xspace}
\newcommand {\MeanNpart} {\mbox{\ensuremath{< \kern-0.15em N_{part} \kern-0.15em >}}}

\newcommand {\sig}       {\ensuremath{S}\xspace}
\newcommand {\expsig}    {\ensuremath{\hat{S}}\xspace}
\newcommand {\prob}      {\ensuremath{P}\xspace}
\newcommand {\prior}     {\ensuremath{C}\xspace}
\newcommand {\prop}      {\ensuremath{F}\xspace}
\newcommand {\atrue}     {\ensuremath{\vec{A}_{\mathrm{true}}}\xspace}
\newcommand {\ameas}     {\ensuremath{\vec{A}_{\mathrm{meas}}}\xspace}
\newcommand {\detresp}   {\ensuremath{R}\xspace}

\newcommand {\pid}       {\ensuremath{\mathrm{\epsilon}_\mathrm{PID}}\xspace}
\newcommand {\nsigma}    {\ensuremath{\mathrm{n_{\sigma}}}\xspace}
\newcommand {\ylab} {\ensuremath{\mathrm{| y_{lab} |}}\xspace}

%
%
\newcommand {\mass}     {\mbox{\rm MeV$\kern-0.15em /\kern-0.12em c^2$}}
\newcommand {\tev}      {\mbox{${\rm TeV}$}\xspace}
\newcommand {\gev}      {\mbox{${\rm GeV}$}\xspace}
\newcommand {\mev}      {\mbox{${\rm MeV}$}\xspace}
\newcommand {\kev}      {\mbox{${\rm keV}$}\xspace}
\newcommand {\tevBoldMath}  {\mbox{${\rm \mathbf{TeV}}$}}
\newcommand {\gevBoldMath}  {\mbox{${\rm \mathbf{GeV}}$}}
\newcommand {\mmom}     {\mbox{\rm MeV$\kern-0.15em /\kern-0.12em c$}}
\newcommand {\gmom}     {\mbox{\rm GeV$\kern-0.15em /\kern-0.12em c$}}
\newcommand {\mmass}    {\mbox{\rm MeV$\kern-0.15em /\kern-0.12em c^2$}}
\newcommand {\gmass}    {\mbox{\rm GeV$\kern-0.15em /\kern-0.12em c^2$}}
\newcommand {\nb}       {\mbox{\rm nb}}
\newcommand {\musec}    {\mbox{$\mu {\rm s}$}}
\newcommand {\nsec}     {\mbox{${\rm ns}$}}
\newcommand {\psec}     {\mbox{${\rm ps}$}}
\newcommand {\fmC}      {\mbox{${\rm fm/c}$}}
\newcommand {\fm}       {\mbox{${\rm fm}$}}
\newcommand {\cm}       {\mbox{${\rm cm}$}}
\newcommand {\mm}       {\mbox{${\rm mm}$}}
\newcommand {\mim}      {\mbox{$ \mu {\rm m}$}}
\newcommand {\cmq}      {\mbox{${\rm cm}^{2}$}}
\newcommand {\mmq}      {\mbox{${\rm mm}^{2}$}}
\newcommand {\dens}     {\mbox{${\rm g}/{\rm cm}^{3}$}}
\newcommand {\lum}      {\, \mbox{${\rm cm}^{-2} {\rm s}^{-1}$}}
\newcommand {\barn}     {\, \mbox{${\rm barn}$}}
\newcommand {\m}        {\, \mbox{${\rm m}$}}
\newcommand {\dg}       {\mbox{$\kern+0.1em ^\circ$}}
\newcommand{\mpp}{\ensuremath{\mathrm{pp}}\xspace}
\newcommand{\rts}{\ensuremath{\sqrt{s}}\xspace}
\newcommand{\GeV}{\ensuremath{\mathrm{GeV}}\xspace}
\newcommand{\TeV}{\ensuremath{\mathrm{TeV}}\xspace}
\newcommand{\gevc}{GeV/\ensuremath{c}\xspace}
\newcommand{\GeVc}{\gevc}
\newcommand{\mevc}{\ensuremath{\mathrm{MeV}/c}\xspace}
\newcommand{\mevcc}{\ensuremath{\mathrm{MeV}/c^{2}}\xspace}
\newcommand{\gevcc}{\ensuremath{\mathrm{GeV}/c^{2}}\xspace}
\newcommand{\pt}{\ensuremath{p_{\rm T}}\xspace}
\newcommand{\kt}{\ensuremath{k_{\rm T}}\xspace}
\newcommand {\lumi}{\mathcal{L}_{\rm int}\xspace}
\newcommand{\nbinv}{\ensuremath{\rm nb^{-1}}}
\newcommand {\ubinv}{\ensuremath{\mu\rm b^{-1}}}
\newcommand {\um}{\ensuremath{\mu\rm m}\xspace}

\newcommand{\lt}{\textless}
\newcommand{\ctau}{\ensuremath{c\tau\xspace}}

%
%

\newcommand{\ePlusMinus}       {\mbox{$\mathrm {e^{\pm}}$}\xspace}
\newcommand{\muPlusMinus}      {\mbox{$\mathrm {\mu^{\pm}}$}\xspace}

\newcommand{\pion}            {\mbox{$\mathrm {\pi}$}\xspace}
\newcommand{\piZero}            {\mbox{$\mathrm {\pi^0}$}\xspace}
\newcommand{\piMinus}           {\ensuremath{\mathrm {\pi^-}}\xspace}
\newcommand{\piPlus}            {\ensuremath{\mathrm {\pi^+}}\xspace}
\newcommand{\piPlusMinus}       {\mbox{$\mathrm {\pi^{\pm}}$}\xspace}

\newcommand{\proton}    {\mbox{$\mathrm {p}$}\xspace}
\newcommand{\pbar}      {\mbox{$\mathrm {\overline{p}}$}\xspace}
\newcommand{\pOuPbar}   {\mbox{$\mathrm {p^{\pm}}$}\xspace}
\newcommand{\DZero}     {\mbox{$\mathrm {D^0}$}\xspace}
\newcommand{\DZerobar}  {\mbox{$\mathrm {\overline{D}^0}$}\xspace}
\newcommand{\Bminus}    {\mbox{$\mathrm {B^-}$}\xspace}
\newcommand{\BZero}     {\mbox{$\mathrm {B^0}$}\xspace}
\newcommand{\BZerobar}  {\mbox{$\mathrm {\overline{B}^0}$}\xspace}

\newcommand{\Dmes}       {\mbox{$\mathrm {D}$}\xspace}
\newcommand{\Lc}         {\mbox{$\mathrm {\Lambda_{c}}$}\xspace}
\newcommand{\Lb}{\ensuremath{\rm {\Lambda_b}}\xspace}
\newcommand{\Xic}         {\mbox{$\mathrm {\Xi_{c}}$}\xspace}
\newcommand{\lambdab}     {\mbox{$\mathrm {\Lambda_{b}^{0}}$}\xspace}
\newcommand{\lambdac}     {\mbox{$\mathrm {\Lambda_{c}^{+}}$}\xspace}
\newcommand{\xicz}        {\mbox{$\mathrm {\Xi_{c}^{0}}$}\xspace}
\newcommand{\xiczp}        {\mbox{$\mathrm {\Xi_{c}^{0,+}}$}\xspace}
\newcommand{\xicp}        {\mbox{$\mathrm {\Xi_{c}^{+}}$}\xspace}
\newcommand{\xib}        {\mbox{$\mathrm {\Xi_{b}}$}\xspace}
\newcommand{\LambdaParticle}        {\mbox{$\mathrm {\Lambda}$}\xspace}

\newcommand{\rmLambdaZ}         {\mbox{$\mathrm {\Lambda}$}\xspace}
\newcommand{\rmAlambdaZ}        {\mbox{$\mathrm {\overline{\Lambda}}$}\xspace}
\newcommand{\rmLambda}          {\mbox{$\mathrm {\Lambda}$}\xspace}
\newcommand{\rmAlambda}         {\mbox{$\mathrm {\overline{\Lambda}}$}\xspace}
\newcommand{\rmLambdas}         {\mbox{$\mathrm {\Lambda \kern-0.2em + \kern-0.2em \overline{\Lambda}}$}\xspace}

\newcommand{\Vzero}             {\mbox{$\mathrm {V^0}$}\xspace}
\newcommand{\Vzerob}             {\mbox{{\bold $\mathrm {V^0}$}}\xspace}
\newcommand{\Kzero}             {\mbox{$\mathrm {K^0}$}\xspace}
\newcommand{\Kzs}               {\ensuremath{\mathrm {K^0_S}}\xspace}
\newcommand{\phimes}            {\ensuremath{\mathrm {\phi}}\xspace}
\newcommand{\Kminus}            {\ensuremath{\mathrm {K^-}}\xspace}
\newcommand{\Kplus}             {\ensuremath{\mathrm {K^+}}\xspace}
\newcommand{\Kstar}             {\mbox{$\mathrm {K^*}$}\xspace}
\newcommand{\Kplusmin}          {\mbox{$\mathrm {K^{\pm}}$}\xspace}
\newcommand{\Jpsi}              {\ensuremath{\rm J/\psi}\xspace}
\newcommand{\DtoKpi}{\ensuremath{\rm D^0\to K^-\pi^+}\xspace}
\newcommand{\DtoKpipi}{\ensuremath{\rm D^+\to K^-\pi^+\pi^+}\xspace}
\newcommand{\DstartoDpi}{\ensuremath{\rm D^{*+}\to D^0\pi^+}\xspace}
\newcommand{\Dzero}{\ensuremath{\mathrm {D^0}}\xspace}
\newcommand{\Dzerobar}{\ensuremath{\mathrm{\overline{D}^0}}\xspace}
\newcommand{\Dstar}{\ensuremath{\rm D^{*+}}\xspace}
\newcommand{\Dplus}{\ensuremath{\rm D^+}\xspace}
\newcommand{\decleng}{\ensuremath{\rm L}_{xyz}}
\newcommand{\Lcminus}{\ensuremath{\rm {\overline{\Lambda}{}_c^-\xspace}}}
\newcommand{\Lcplus}{\ensuremath{\rm {\Lambda_c^+}\xspace}}
\newcommand{\Lbzero}{\ensuremath{\rm {\Lambda_b^0}}\xspace}
\newcommand{\LctopKpi}{\ensuremath{\rm \Lambda_{c}^{+}\to p K^-\pi^+}\xspace}
\newcommand{\LbtoLc}{\ensuremath{\rm \Lambda_{b}^{0}\to \Lc + \rm{X}}\xspace}
\newcommand{\LctopKzS}{\ensuremath{\rm \Lambda_{c}^{+}\to p K^{0}_{S}}\xspace}
\newcommand{\LctoenuLambda}{\ensuremath{\rm \Lambda_{c}^{+}\to e^{+} \nu_{e} \Lambda}\xspace}
\newcommand{\cosP}{\ensuremath{\rm cos_{\Theta_{pointing}}}\xspace}
\newcommand{\KzStopippim}{\ensuremath{\rm K^{0}_{S}\to \pi^{+} \pi^{-}}\xspace}
\newcommand{\Lambdatoppim}{\ensuremath{\rm \Lambda \to p \pi^{-}}\xspace}
\newcommand{\nue}{$\nu_e$}
\newcommand{\DtopiKzs}{\ensuremath{\rm D^+\to \pi^+ K^{0}_{S}}\xspace}
\newcommand{\DstoKKzs}{\ensuremath{\rm D_s^+\to K^+ K^{0}_{S}}\xspace}

\newcommand{\ptLc}{\ensuremath{p_{\rm T, \Lambda_c}}\xspace}
\newcommand{\ptpion}{\ensuremath{p_{\rm T, \pi}}\xspace}
\newcommand{\ptK}{\ensuremath{p_{\rm T, K}}\xspace}
\newcommand{\ptproton}{\ensuremath{p_{\rm T, \proton}}\xspace}



\begin{titlepage}
\PHyear{2017}  
\PHnumber{339}          
\PHdate{22 December}              
\title{\Lcplus~production in pp collisions at \sqrts = 7~TeV and in \pPb collisions at \sqrtsNN = 5.02~TeV}
\ShortTitle{\Lcplus~ production in ALICE}   
%
\Collaboration{ALICE Collaboration%
         \thanks{See Appendix~\ref{app:collab} for the list of collaboration members}}
\ShortAuthor{ALICE Collaboration}      
\begin{abstract}
The \pt-differential production cross section of prompt \Lcplus~charmed baryons was measured with the \-ALICE detector at the 
Large Hadron Collider (LHC) in \pp collisions at \sqrts $= 7$ \tev and in \pPb collisions at \sqrtsNN $= 5.02$ \tev at midrapidity.
The \Lcplus~ and \Lcminus~ were reconstructed in the hadronic decay modes \LctopKpi, \LctopKzS and in the semileptonic
channel \LctoenuLambda (and charge conjugates).
The measured values of the \Lcplus/\Dzero ratio, which is sensitive to the c-quark hadronisation mechanism, and in particular to the production of baryons,
are presented and are larger than those measured previously in different colliding systems, centre-of-mass energies, rapidity and \pt intervals,
where the \Lcplus~production process may differ. 
The results
are compared with the expectations obtained from perturbative Quantum Chromodynamics calculations and Monte Carlo event generators. Neither perturbative QCD calculations nor Monte Carlo models reproduce the data, indicating that the fragmentation of heavy-flavour baryons is not well understood.
The first measurement at the LHC of the \Lcplus~nuclear modification factor, \RpPb, is also presented.
The \RpPb is found to be consistent with unity and with that of D mesons
within the uncertainties, and consistent with a theoretical calculation that includes cold nuclear matter effects and a calculation that includes charm quark interactions with a deconfined medium.

\end{abstract}
\end{titlepage}
\setcounter{page}{2}

\cleardoublepage
\clearpage

\section{Introduction}
\label{sec: Introduction}

The study of charm production at the Large Hadron Collider (LHC) is an  important tool to test predictions obtained from perturbative 
Quantum Chromodynamics (pQCD) calculations for proton--proton (\pp) collisions.
These calculations are based on the factorisation approach that describes heavy-flavour production as a convolution of the parton distribution
functions, the parton hard-scattering cross section and the fragmentation function. The cross section for
heavy-flavour hadron production
can be obtained from perturbative calculations at next-to-leading order with next-to-leading-log resummation, like the General-Mass Variable-Flavour-Number Scheme (GM-VFNS~\cite{Kniehl:2005mk,Kniehl:2012ti}) and Fixed-Order Next-to-Leading-Log (FONLL~\cite{Cacciari:1998it,Cacciari:2012ny}) approaches.
No predictions are, however, available for baryons in the latter approach due to lack of knowledge of the fragmentation function of charm quarks into baryons.
Cross section calculations are available also with the \kt factorisation framework~\cite{Maciula:2013wg}. These theoretical calculations generally describe within uncertainties the measurements at the LHC, with the central predictions for beauty production lying closer to data than the central predictions for charm production~\cite{Andronic:2015wma}. The measured transverse momentum differential cross section of charm mesons lies in the upper part of the FONLL uncertainty band and is systematically below the central value of GM-VFNS predictions~\cite{Acharya:2017jgo}. Cross sections for charm production are also available in general-purpose Monte Carlo generators such as {\sc pythia}~\cite{Sjostrand:2007gs}. The hard process amplitude is calculated with leading order (LO) accuracy and, via parton showers, effective LO+LL accuracy is provided. Next-to-Leading-Order (NLO) Monte Carlo generators were developed by matching event generators, calculating the hard scattering with NLO accuracy, as in {\sc powheg}~\cite{Frixione:2007nw}, with parton showers as in {\sc pythia}.

In pQCD calculations, the hadronisation process is modeled via a fragmentation function, which para\-metrises the fraction of the quark energy transferred to the produced hadron, and by the fragmentation fractions, which account for the probability of a heavy quark to hadronise into a particular hadron species. Fragmentation functions are tuned on electron--positron data under the assumption that they are universal. Similarly, the fragmentation fractions were usually assumed to be the same in different collision systems.
Among other observables, the relative production of baryons and mesons (``baryon-to-meson ratio'') is particularly sensitive to the fragmentation process. A study of the \lambdab baryon to \Bminus and \BZerobar meson production by LHCb~\cite{Aaij:2011jp} reported a transverse momentum (\pT) dependence of that ratio, interpreted as evidence of non-universality of fragmentation fractions in the beauty sector~\cite{Amhis:2016xyh,Olive:2016xmv}. In Monte Carlo generators, hadronisation is implemented via formation of strings as in {\sc pythia}, via ropes~\cite{Biro:1984cf} as in {\sc dipsy}~\cite{Flensburg:2011kk} or via clusters as in {\sc herwig}~\cite{Bahr:2008pv}. In hadron--hadron collisions at LHC energies, multi-parton interactions and coherence effects between multiple partonic interactions may affect the hadronisation processes.   Within the existing {\sc pythia\small8} framework a better agreement with measurements by CMS~\cite{Khachatryan:2011tm} of the \rmLambda/\Kzs ratio was obtained in~\cite{Christiansen:2015yqa} introducing additional colour reconnection mechanisms that play a role in \pp collisions and are instead expected to be highly suppressed in electron--positron collisions at LEP. For the {\sc dipsy} event generator in~\cite{Bierlich:2015rha} an approach was tested where strings from independent interactions can be close in space and form colour ropes, expected to yield more baryons than independent strings.
Therefore, the measurement of the \Lcplus~production cross section in \pp collisions allows one to test these expectations at LHC energies with charmed baryons and mesons.

Furthermore, the study of charmed-baryon production could play an important role in the investigation of the state of strongly-interacting matter at very high temperatures
and densities realised in heavy-ion collisions, known as the Quark-Gluon Plasma (QGP)~\cite{BraunMunzinger:2007zz}. Measurements of open heavy-flavour production in this environment allow for the study of the interaction of heavy quarks with the medium constituents and the characterisation of the properties of the plasma state~\cite{Rapp:2009my}.
The interaction with the medium constituents could modify the hadronisation: a significant fraction of low and intermediate-momentum charm and beauty quarks could hadronise via recombination (coalescence) with other quarks from the medium~\cite{Greco:2003vf, Oh:2009zj}.
Models including coalescence predict an enhanced baryon-to-meson ratio at low and intermediate \pT relative to that observed in \pp collisions where hadronisation can be described by string-fragmentation models~\cite{Sjostrand:2007gs}. In addition, the possible existence of
light diquark bound states in the QGP could further enhance the \Lcplus/\Dzero
ratio in coalescence models~\cite{Lee:2007wr}. An enhancement of the \pT-integrated \Lcplus/\Dzero ratio in presence of a QGP is also predicted by statistical hadronisation models\cite{Kuznetsova:2006bh}, where the relative abundance of hadrons depends only on their masses and on the freeze-out temperature of the medium created in the collision.
Recently, such an enhancement of the \Lcplus/\Dzero ratio was preliminarily reported by STAR in Au--Au collisions at \sqrtsNN =~200 GeV in the $3<\pT<6$~\gevc interval~\cite{Xie:2017jcq}. A measurement of prompt \Lcplus~production at the LHC in \pp collisions is needed as a baseline reference for these studies.

For the intepretation of the results in nucleus--nucleus collisions, the measurement in proton--nucleus collisions is also crucial. In such a system cold-nuclear-matter (CNM) effects can affect the production of charm hadrons:
their assessment is needed to disentangle them from the effects related to the formation of the QGP (hot-medium effects). In the initial state, the Parton
Distribution Functions (PDFs) are modified in bound nucleons compared to free nucleons. The nuclear shadowing at low transverse momentum can decrease, among other effects, the production cross section of open charm~\cite{Arneodo:1992wf}.
Moreover, the multiple scattering of partons in the nucleus before or after the hard scattering can affect the momentum distributions of the produced hadrons, especially at low \pT (\pT \textless~2 \gevc). In addition to initial-state effects, final-state effects may also be responsible for the modification of particle yields and transverse-momentum distributions in proton--nucleus collisions as compared to \pp interactions.
Nuclear effects can be investigated measuring the nuclear modification factor \RpPb, defined as the ratio of the cross section in \pPb collisions to that in \pp interactions scaled by the mass number of the Pb nucleus.
A recent measurement~\cite{Adam:2016ich,ALICE:2017pps} of D-meson production in \pPb collisions showed that, within uncertainties,
\RpPb is compatible with unity, indicating that initial and final-state effects are either small or that they compensate each other. Several other observations in \pPb collisions, such as the presence of di-hadron azimuthal correlations at large rapidity differences~\cite{CMS:2012qk,Abelev:2012ola,Abelev:2013wsa,Aad:2012gla,Adam:2015bka}, the evolution of the average \pT at central rapidity of identified hadrons with multiplicity~\cite{Abelev:2013haa, Chatrchyan:2013eya} and the increased strangeness yield with increasing multiplicity~\cite{Adam:2015vsf} qualitatively resemble observations in \PbPb collisions. This suggests the possible formation of a hot deconfined medium also in \pPb collisions that, in turn, can affect the propagation and hadronisation of heavy quarks, modifying the momentum distribution of the charmed hadrons with respect to that expected from \pp collisions, hence inducing a deviation of \RpPb from unity~\cite{Beraudo:2015wsd,Xu:2015iha}.

At high energies, \Lcplus~production has been studied at electron--positron colliders (at the Z-resonance with LEP~\cite{Alexander:1996wy,Barate:1999bg,Abreu:1999vw}, and at B factories~\cite{Albrecht:1988an,Avery:1990bc,Seuster:2005tr,Aubert:2006cp}), in several fixed target experiments including neutrino--proton~\cite{KayisTopaksu:2003mh}, hadron--nucleon~\cite{Garcia:2001xj} and photon--nucleon~\cite{Link:2002ge} interactions and at electron--proton colliders (in photoproduction~\cite{Chekanov:2005mm,Abramowicz:2013eja}, and via deep inelastic scattering~\cite{Abramowicz:2010aa}). At the LHC, a measurement of \Lcplus-baryon production at forward rapidity was reported by the LHCb Collaboration~\cite{Aaij:2013mga} in \pp collisions at \sqrts = 7 \tev in the rapidity ($y$) range $2.0<y<4.5$. Here and in the following, $y$ is defined in the centre-of-mass system of the collision. A preliminary result in \pPb collisions at \sqrtsNN = 5.02 \tev has also been presented recently by LHCb~\cite{LHCb:2017rvh}. Previous measurements at hadron--hadron colliders \cite{Drijard:1979vd,Basile:1981wg,AguilarBenitez:1987cw} are at a much lower centre-of-mass energy (\sqrts = O(100) \gev). 

In this paper, we present the measurement of the production cross section of the prompt charmed baryon \Lcplus (udc) and its charge conjugate (c.c.). Hereafter with \Lc we will refer indistinctly to both, and all mentioned decay channels refer also to their charge conjugate.  The contribution from beauty feed-down to the measured \Lc yields was subtracted by using pQCD calculations of the beauty-hadron cross section together with the acceptance and efficiency values extracted from simulation.
The cross section was measured with the ALICE detector~\cite{Aamodt:2008zz} in \pp collisions at $\sqrts = 7~\tev$ in the transverse momentum and rapidity intervals $1<\pT<8$ \gevc and $|y| < 0.5$ and in \pPb collisions at \sqrtsNN = 5.02 \tev in $2<\pT<12$ \gevc  and $-0.96 < y < 0.04$.

Due to the short lifetime of the \Lc baryons (c$\tau$ = 60 $\mu$m~\cite{Olive:2016xmv}) and the statistical limitation of the data sample considered,
the reconstruction of \Lc decays was particularly challenging. Three decay channels of the \Lc were therefore studied, two hadronic channels (\LctopKpi and \LctopKzS), and a semileptonic one (\LctoenuLambda). Furthermore, se\-veral different independent analysis strategies were developed, including the use of a Bayesian approach for particle identification~\cite{Adam:2016acv} and a Multivariate Analysis (MVA)~\cite{Hocker:2007ht}. These developments build on top of the tools and strategies used in previous ALICE analyses of D-meson hadronic decays \cite{ALICE:2011aa,Abelev:2012tca,Abelev:2012vra,Adam:2016ich,Acharya:2017jgo,ALICE:2017pps} and of the \Xic-baryon semileptonic decay~\cite{Xic}.  After a description of the detector and the data samples in \secref{sec: Data samples and experiment}, we detail the different analyses
and methods used for the various decay channels and collision systems in \secref{sec: Analysis overview and methods}. The efficiency corrections applied and the treatment of the feed-down correction are described in \secref{sec: Corrections}. The evaluation of the systematic uncertainties is presented in \secref{sec: Systematics}. Finally, the results are presented and discussed in \secref{sec: Results}. Here, the cross section measured in \pp collisions and the \Lcplus/\Dzero production ratio are compared with pQCD calculations and predictions from event generators as well as with existing measurements in different collision systems and rapidity intervals. The cross section obtained in \pPb collisions is compared with the \pp results, and the first measurement of the \Lcplus~nuclear modification factor in \pPb collisions, \RpPb, is presented.

\section{Experimental setup and data samples}
\label{sec: Data samples and experiment}
A comprehensive description of the ALICE apparatus and its performance 
can be found in \cite{Aamodt:2008zz, Abelev:2014ffa}. In this section, the detectors 
used for the analyses discussed in this paper are described. 
\Lc baryons were measured by reconstructing their decay products in the pseudorapidity interval $|\eta|<0.8$ relying on the
tracking and particle identification (PID) capabilities of the central-barrel detectors, which are located in a solenoid magnet providing a $B$ = 0.5 T field, parallel to the beam direction ($z$-axis in the ALICE reference frame). In particular, the Inner Tracking System (ITS) and the Time Projection Chamber (TPC) were utilised for track reconstruction, while PID was performed based on the information from the TPC and the Time-Of-Flight detector (TOF).

From the innermost radius of 3.9 cm (distance from the centre of the beam vacuum tube) to the outermost radius of 43.0 cm, the ITS cylinder includes two layers of Silicon Pixel Detector (SPD), two Silicon Drift Detector layers, and two Silicon Strip Detector layers. The different ITS detectors have full azimuth but different pseudorapidity coverage, with a common $|\eta| < 0.9$ acceptance. 
The spatial precision of the ITS detector,
its vicinity to the beam pipe, and its very low material budget \cite{Aamodt:2010aa}
allow for a precise determination of the track impact parameter (i.e. the distance of closest approach of the track to the primary vertex) in the transverse plane, for which a resolution better than 75~$\mu$m  is achieved for tracks with \pt $> 1$ \gevc ~\cite{Aamodt:2010aa}. 

The TPC is the main tracking detector of the experiment and surrounds the ITS with an active radial range from 85 cm to 250 cm and with full azimuthal coverage in the pseudorapidity interval $|\eta| < 0.9$. It provides up to 159 space points to reconstruct the particle trajectory and determine its momentum. Additionally, it provides particle identification via the measurement of the specific energy loss, \dEdx. The TOF (an array of 1593 Multi-gap Resistive Plate Chambers) completes the set of detectors used for PID in the analyses presented in this paper. It is located at a radial distance of about 3.8 m, covering full azimuth in the pseudorapidity interval $|\eta| < 0.9$.
The particle arrival time at the detector is determined with a resolution of about 80 ps. The T0 consists of two arrays of Cherenkov counters, located on both sides of the interaction point at +350 cm and $-$70 cm from the nominal vertex position along the beam line. The time resolution of the T0 in pp and \pPb collisions is about 50 ps for the events in which the 
measurement is made on both sides~\cite{Adam:2016ilk}.
The event time of the collision is obtained on an event-by-event basis either using the TOF detector, or the T0 detector, or a combination of the two \cite{Adam:2016ilk}.
  
The results presented in this paper were obtained from the analysis of the \run{1} data collected by ALICE in \pp collisions at \sqrts = 7 TeV in 2010 and in \pPb collisions at \sqrtsNN = 5.02 TeV from the 2013 data taking campaign. During the \pPb run, the beam energies were 4 TeV for protons
and 1.59 TeV per nucleon for lead nuclei. With this beam configuration, the proton--nucleon centre-of-mass system moves in rapidity by $\Delta y$ = 0.465 in the direction of the proton beam.

The V0 detector, used for trigger and event selection, consists of two scintillator arrays, called V0A and V0C, covering the full azimuth in the pseudorapidity intervals $2.8 < \eta < 5.1$ and $-3.7 < \eta < -1.7$, respectively.   
The analyses used events recorded with a minimum bias (MB) trigger, which was based on the signals from the V0 and SPD detectors.
At least one hit in either of the two scintillator arrays of the V0, or at least one hit in the SPD (pseudorapidity coverage of $|\eta| < 2$ and $|\eta| < 1.4$ for the inner and the outer layers, respectively) was required by the MB-trigger condition during the \pp data taking, while in \pPb the requirement was based on coincident hits in both V0A and V0C.
The events were further selected offline using the SPD, V0 and Zero Degree Calorimeter (ZDC) 
information in order to remove background from beam-gas collisions, and from the machine as described in 
\cite{Aamodt:2009aa,ALICE:2012xs}. In the analysed sample, events with more than one interaction (pile-up) were removed according to the vertex information reconstructed from the hits in the SPD detector. To  maximise the ITS acceptance, only events with a $z$-coordinate of the reconstructed vertex position within 10 cm from the nominal interaction point were used.  
With these requirements, approximately 300 and 370 million MB triggered events were analysed for the \pp hadronic and semileptonic channels, respectively,
corresponding to an integrated luminosity of $\lumi$ = 4.8 and 5.9 \nbinv~with an uncertainty of $\pm$ 3.5\%~\cite{Abelev:2012sea},
while approximately 100 million MB triggered events were selected for the \pPb analyses, corresponding to $\lumi$ = 47.8 \ubinv ($\pm$ 3.7\%~\cite{Abelev:2014epa}).

\section{\Lc analysis overview and methods}
\label{sec: Analysis overview and methods}

The measurement of \Lc production was performed
by reconstructing three decay modes:
\LctopKpi with branching ratio (BR) equal to (6.35 $\pm$ 0.33)$\%$,
\LctopKzS with BR = (1.58 $\pm$ 0.08)$\%$ and \KzStopippim with BR = (69.20 $\pm$ 0.05)$\%$, and \LctoenuLambda with BR = (3.6 $\pm$ 0.4)$\%$
and \Lambdatoppim with BR = (63.9 $\pm$ 0.5)$\%$~\cite{Olive:2016xmv}.
The hadronic decays were fully reconstructed while the semileptonic decay was partially reconstructed because the neutrino is not detectable with the ALICE setup.
The analysis strategy for the extraction of the \Lc signals from the large combinatorial background was based on the reconstruction of charged tracks with the central-barrel detectors, on the V-shaped neutral decay topology reconstruction (V$^0$) of \Kzs and $\Lambda$, on kinematical and geometrical selections, and on the use of PID on the decay tracks.

These analyses cannot fully benefit of the reconstruction and selection of
secondary vertex topologies due to the comparable resolution of the ITS on the track impact parameter and the mean decay length of the \Lc.
The use of PID techniques is therefore fundamental to reduce the large combinatorial background.
The identification of pions, kaons, protons, and electrons used for the \Lc analyses in all the considered decay channels and for both colliding
systems was based on the information from the specific energy loss \dEdx in the TPC detector
and on the time of flight measured with the TOF detector.
For some of the results presented here, MVA techniques were applied additionally to the selection procedure based on classical cuts and called ``standard'' (STD) in the following. 
Finally, the \Lc raw yield was extracted with an invariant mass analysis for the hadronic decay modes or, in the semileptonic analysis, 
by counting the candidates with the correct combination of particle species and charge sign (i.e. e$^+ \Lambda$ and e$^- \overline{\Lambda}$), 
indicated as ``right sign'' in the following,
after subtracting the background estimated from ``wrong sign'' pairs (i.e. e$^- \Lambda$ and e$^+ \overline{\Lambda}$).
Table~\ref{tab:strategies} summarises the various analysis methods.

Simulations were used in the analyses to determine the geometrical acceptance, the efficiencies of track reconstruction and \Lc selection, and the line shape of the \Lc invariant-mass peak.  The event generator used to simulate \pp
collisions was {\sc pythia \small6.4.21}~\cite{Sjostrand:2006za}
with the Perugia-0 tune~\cite{Skands:2009zm}. For \pPb collisions, {\sc pythia} events containing a c$\overline{\rm{c}}$ or b$\overline{\rm{b}}$ pair were merged with events simulated with the {\sc hijing \small1.36} event generator~\cite{Wang:1991hta} to obtain a better description of the multiplicity distribution observed in data.
The generated particles were transported through the ALICE detector
using the {\sc geant\small3} package~\cite{Brun:1994aa}.

\begin{table}
\begin{center}
\def\arraystretch{1.2}\tabcolsep=6pt   
\begin{tabular}{lcccc}
\toprule
& & & \multicolumn{2}{c}{Strategy}  \\
\cmidrule{4-5} 
Decay channel & System & \sqrtsNN (TeV) & Method & PID \\
 \midrule
\LctopKpi                                   & \multirow{3}{*}{\pp}  & \multirow{3}{*}{7}             & STD & Bayes \\
\LctopKzS                                  &                                &                                           & STD & n$_{\sigma}$ \\ 
\LctoenuLambda                       &                                 &                                          & Pair combination & n$_{\sigma}$ \\  
\midrule
\multirow{2}{*}{\LctopKpi}          & \multirow{4}{*}{\pPb} & \multirow{4}{*}{5.02}      & STD & Bayes \\
                                                  &                                   &                                       & MVA & n$_{\sigma}$,Bayes \\[0.2cm]
\multirow{2}{*}{\LctopKzS}         &                                  &                                        & STD & n$_{\sigma}$ \\ 
                                                  &                                   &                                       &  MVA & n$_{\sigma}$,Bayes\\
\midrule
\end{tabular}
\caption{\Lc decay channels studied and analysis methods presented in this paper.}
\label{tab:strategies}
\end{center}
\end{table}
            
For all the analyses, the lower limit of the  \Lc \pt interval in which the signal could be extracted was imposed by the 
large combinatorial background, which could not be reduced enough with the applied selections.
The upper limit was imposed by the limited size of the analysed data sample. 
This section gives an overview of the analysis methods, 
with Sec.~\ref{sec:HadronicDecayModes} dedicated to the \Lc hadronic decay modes
and Sec.~\ref{sec:SemileptonicDecayMode} to the semileptonic channel.

\subsection{Hadronic decay modes}
\label{sec:HadronicDecayModes}

The \LctopKpi candidates were built from triplets of reconstructed tracks with proper charge-sign combination.
The \LctopKzS candidates were constructed by combining a reconstructed track (the bachelor) with a \Kzs candidate.
The \Lc and \Kzs
candidates were formed by combining reconstructed tracks having $\lvert\eta\lvert\lt0.8$ and at least 70 associated space points in the TPC.
Additionally, for the bachelor and the tracks used to form \LctopKpi candidates, at least one hit in either of the two SPD layers was required.
The \Kzs candidates were identified by applying selections on characteristics of their decay tracks (\pt $> 0.1$ \gevc,
a minimum transverse impact parameter to the primary vertex, $d_0$, of 0.05 cm and a maximum distance of closest approach between the daughters tracks of 1.5 cm)
and of their weak decay topology (a minimum transverse decay radius of 0.2 cm and a minimum cosine of the V$^0$ pointing angle to the primary vertex of 0.99).
The invariant mass of the $\pi^+\pi^-$ pair was required to be compatible with the PDG mass of the \Kzs
within 1 or 2 $\sigma$ depending on the \pt interval and the collision system.
To further improve the \Kzs signal purity, especially at lower $\pt$, veto selections on $\Lambda$, $\overline{\Lambda}$ and $\gamma$ PDG masses
were applied to the invariant masses calculated with the p$\pi^-$, $\overline{\mathrm{p}}\pi^+$ and e$^+$e$^-$ hypotheses for the daughter tracks, respectively.

For both decay channels, cuts on kinematical and geometrical variables were also applied after a tuning procedure in each \pt interval.
The kinematical variables include the \pt of the daughter tracks and the \pt of the \Kzs in the \LctopKzS analysis. 
In the \LctopKpi analysis, the geometrical variables include the separation
between the interaction point and the points of closest approach of the opposite-sign track pairs,
the separation between the reconstructed \Lc-decay vertex and the interaction point (decay length),
the distance of closest approach of the three pairs of tracks,
the quadratic sum of the minimum distances of the tracks from the reconstructed \Lc-decay vertex, and the \Lc pointing angle to the primary vertex.
In the \LctopKzS analysis, the geometrical variables include the upper cuts on the $d_0$ of the bachelor and \Kzs
(applied to remove secondary tracks originating very far from the interaction point).
For both decay channels the cuts
were tuned on Monte Carlo samples for each analysis to achieve a high statistical significance in each \pt interval.

After the selection, the acceptance in rapidity for \Lc baryons drops steeply to zero for $|y_{\rm lab}|>0.5$ at low \pt and for $\lvert y_{\rm lab}\lvert>0.8$ 
at $\pt>5$~\gevc, where $y_{\rm lab}$ is the rapidity in the laboratory frame. A \pt-dependent fiducial acceptance cut was therefore applied on the 
\Lc rapidity, $|y_{\rm lab}|<y_{\rm fid}(\pT)$ with $y_{\rm fid}(\pT)$ increasing from 0.5 to 0.8 in the interval $0 < \pT < 5$ \gevc, and $y_{\rm fid}=0.8$ for \pT $> 5$ \gevc, as described in~\cite{Acharya:2017jgo}.

\begin{figure}[t!]
\centering
\includegraphics[width=1.0\textwidth]{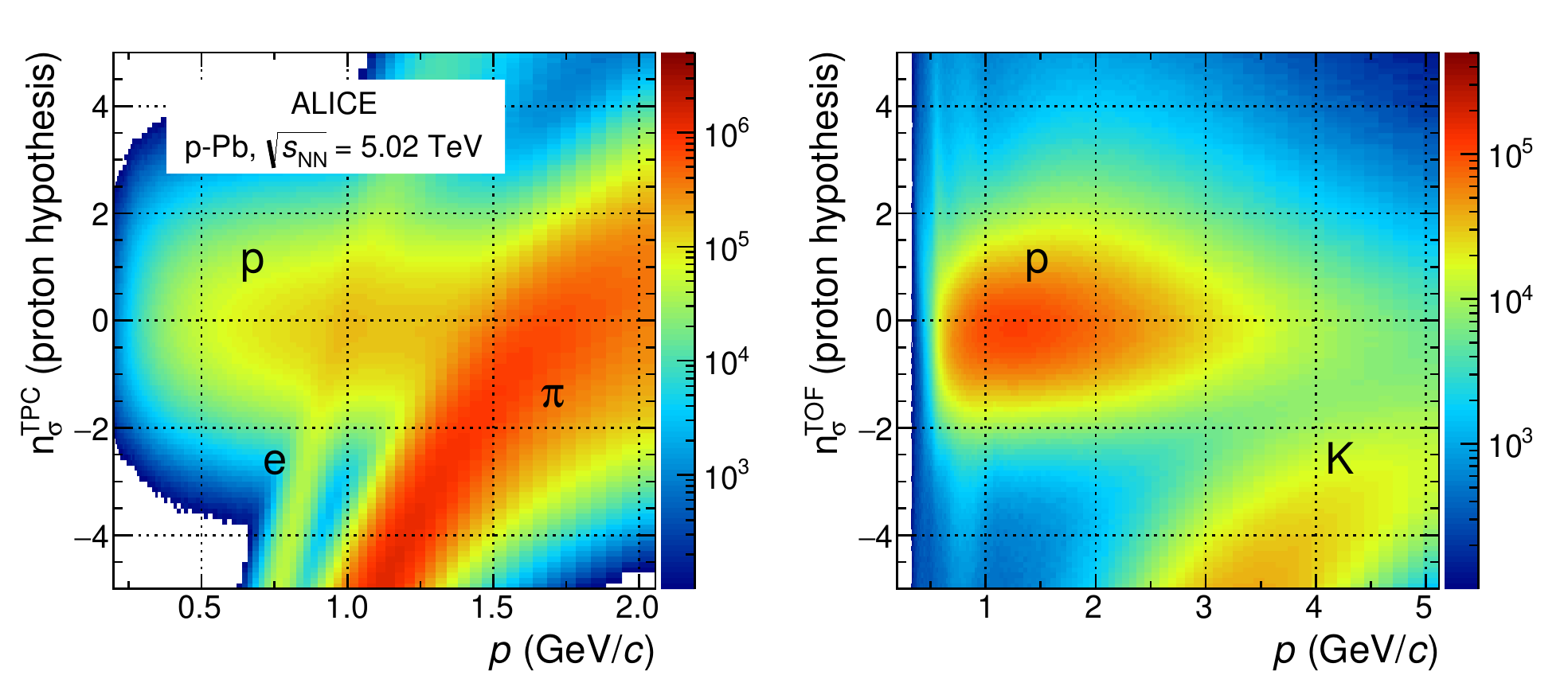} 
\caption{Proton identification with TPC (left) and TOF (right) in \pPb collisions. The discriminating PID variable $n_\sigma$ (see text for details) is shown as a function of the momentum $p$ of the particle. The $n_\sigma$ variable is computed assuming the proton hypothesis. The contributions from electrons and pions in the TPC and from kaons in the TOF are indicated.}
\label{fig:PIDperformanceData}
\end{figure}

The identification of the proton in the \LctopKzS analysis was based on the \dEdx and time-of-flight information, using as a PID-discriminating variable the difference between the measured signal
and that expected under the proton mass hypothesis divided by the detector resolution ($n_\sigma$), as detailed in~\cite{Adam:2016acv}.
Figure~\ref{fig:PIDperformanceData} shows an example of the $n_\sigma$ distributions relative to the proton hypothesis as a function of 
momentum for TOF and TPC signals in \pPb collisions.
To reduce the pion and kaon contamination, for tracks with momentum $p<1$ \GeVc,
a $|n_\sigma|<2$ selection with respect to the proton hypothesis was applied on the TPC \dEdx.
For $p>1$ \GeVc, in order to improve the signal over background ratio, the presence of the TOF signal was requested
and the track was required to be within $ |n_\sigma|<3$ of the expected proton TOF signal,
without any further selections based on TPC information. In this momentum region, tracks missing the TOF information were discarded.
In the \pPb analysis it was further required that the track should be within $|n_\sigma|<3$ of the expected TPC signal.

In the \LctopKpi analysis, where a larger combinatorial background is present,
the Bayesian PID method~\cite{Adam:2016acv} was adopted to increase the purity of the signal.
In this method, the signals from the TOF and TPC are combined
constructing a conditional probability that a given track corresponds to a given hadron species (p, K or $\pi$) based on
a set of measurements in the two detectors. The computation of the Bayesian probability entails the use of priors, that are
evaluated with data-driven techniques. 
This approach provides a smoother increase of the PID efficiency with \pT than the one observed with the $n_\sigma$-cut approach and it makes
the best possible use of the combined information coming from the two detectors.
To each of the three \Lc decay tracks, a single mass hypothesis was assigned, corresponding to the hadron species (p, K, $\pi$) for which the Bayesian probability was found to be the maximum. Candidates were rejected if the daughter-track species and charge sign did not match
with a $\mathrm{pK^-\pi^+}$ (or charge conjugate) final state.
This corresponds to the ``maximum probability'' strategy discussed in~\cite{Adam:2016acv} that was, for example,
successfully validated in reproducing the published results~\cite{ALICE:2011aa} for the \Dzero$\rightarrow$ K$^-\pi^+$ production cross section,
which were obtained with a PID strategy based on a $|n_\sigma|<3$ selection.

In addition to the STD analyses for the study of the hadronic decay modes in \pPb collisions, 
a further analysis was carried out that relies on a multivariate selection to separate the background from the signal, based on Boosted Decision Trees (BDT)~\cite{Hocker:2007ht}. 
This approach will be indicated as MVA in the following. To train the algorithm, the signal sample was built using $\Lc$ particles from Monte Carlo simulations. For the background sample, as detailed later, both real and simulated events were used. This training sample was used to determine a mapping function,
which describes a decision boundary, optimised in order to maximise signal/background separation. The learned mapping function was then applied to a real data sample, in which the type of candidate is unknown. A cut on this decision boundary 
aims to reject background candidates while keeping signal candidates.

Prior to the BDT decision, for both decay channels, PID selections were applied.
For the $\LctopKpi$ analysis a $|n_\sigma|<3$ cut was applied on the compatibility with the expected \dEdx and time-of-flight values.
For proton and kaon identification, tracks without a TOF signal were identified using only the TPC, and tracks with incompatible TPC and TOF identifications were assigned the identity given by the TOF. For pion identification only the TPC was used. In the case of the \LctopKzS analysis, a $|n_\sigma|<3$ compatibility cut was applied on the TPC and TOF, when available, for the bachelor track. For this analysis, an additional cut in the Armenteros-Podolanski space~\cite{Armenteros} was also applied in order to reject $\Lambda$ decays.

Independent BDTs were trained per \pt interval and applied on the \pPb data sample. The BDTs were trained using signal samples consisting of $\Lc$ decays from simulated events, required to have at least one \Lc
per event decaying to either a $\mathrm{pK\pi}$ or $\mathrm{p\Kzs}$ final state, and including a detailed description of the de\-tector response, the geometry of the apparatus and the conditions of the luminous region. The background sample was taken
from the sidebands of the candidate invariant-mass distribution in the data ($\mathrm{pK\pi}$ analysis), or from the simulated events ($\mathrm{p\Kzs}$ analysis), 
and it was verified that swapping the simulated/real background sample does not change the result of the trained BDT.

\begin{figure}[h!]
\centering
\includegraphics[width=1.0\textwidth]{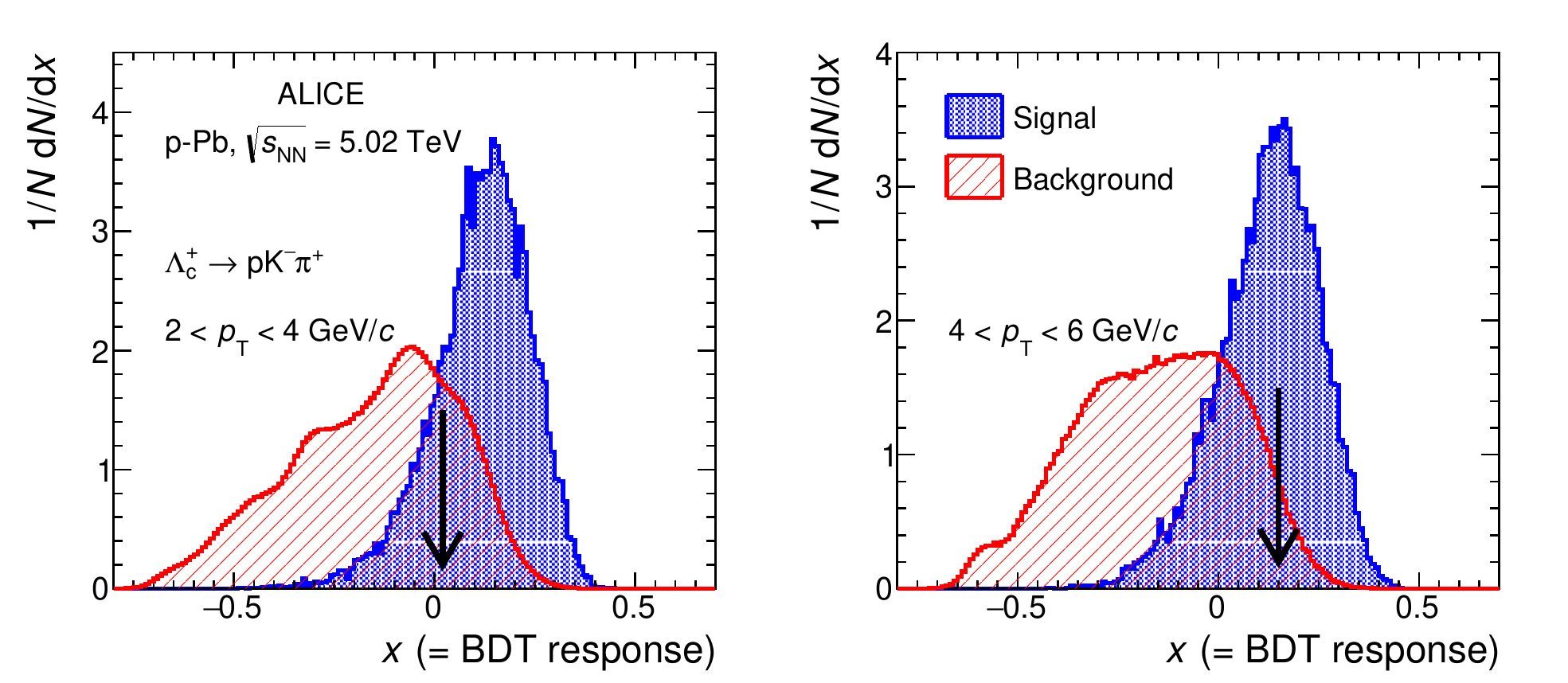}
\caption{Normalised distribution of the BDT responses of the \Lc candidates for Monte Carlo signal (blue area) and background (red shaded area) in two \pt intervals for the \LctopKpi decay channel in \pPb collisions where the MVA method was used. The arrows correspond to the applied cuts.}
\label{fig:BDTResponse}
\end{figure}

For the  \LctopKzS analysis the variables related to the decay topology that were used in the multivariate analysis include the \pt of the bachelor track, the $d_0$ of the bachelor track, the V$^0$ invariant mass under the hypothesis that the daughters are a $\pi^+\pi^-$ pair, the $d_0$ and the lifetime of the V$^0$. For the \LctopKpi analysis the variables related to the decay topology that were used in the multivariate analysis include all variables used in the \LctopKpi STD analysis, as well as the projection of the decay length in the transverse plane normalised by its error. PID variables were also used in both analyses, namely the Bayesian probabilities that each track is correctly identified as either a proton, a kaon, or a pion
for the $\mathrm{pK\pi}$ analysis, and the Bayesian probability that the bachelor track is a proton for the $\mathrm{p\Kzs}$ analysis.
Fi\-gure~\ref{fig:BDTResponse} shows examples of the BDT response in the two lowest \pt intervals for the analysis of the \LctopKpi decay channel.

The raw signal yields 
were extracted by fitting the invariant mass distributions of the \Lc candidates passing the selections outlined above, for every \pt interval under study.
The fitting function consists of a Gaussian describing the signal, whose width was fixed to the value obtained in the simulation, and a polynomial of second order or a linear function (with the choice 
depending on the $\pt$ interval) to describe the background. 

\begin{figure}[ht!]
\centering
\includegraphics[width=1.0\textwidth]{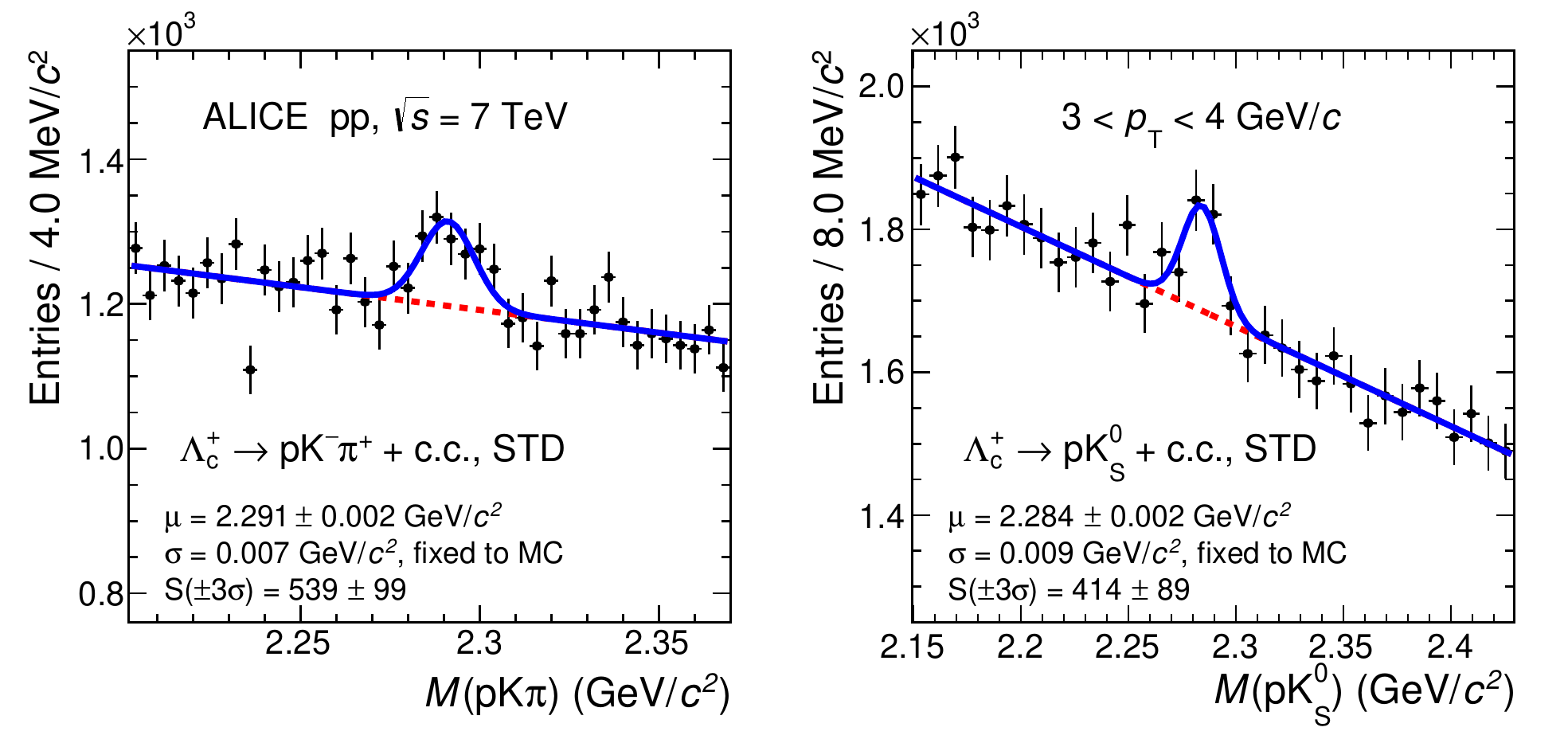}
\caption{Invariant-mass distribution of $\Lcplus$ candidates (and charge conjugates) for $3 <$ \pt $< 4$ \gevc in pp collisions at \sqrts = 7~\tev. The dashed lines represent the fit to the background while the solid lines represent the total fit function. Left: $\LctopKpi$ STD analysis, right: $\LctopKzS$ STD analysis.}
\label{fig:InvMasspp}
\end{figure}

\begin{figure}[htb!]
\centering
\includegraphics[width=1.0\textwidth]{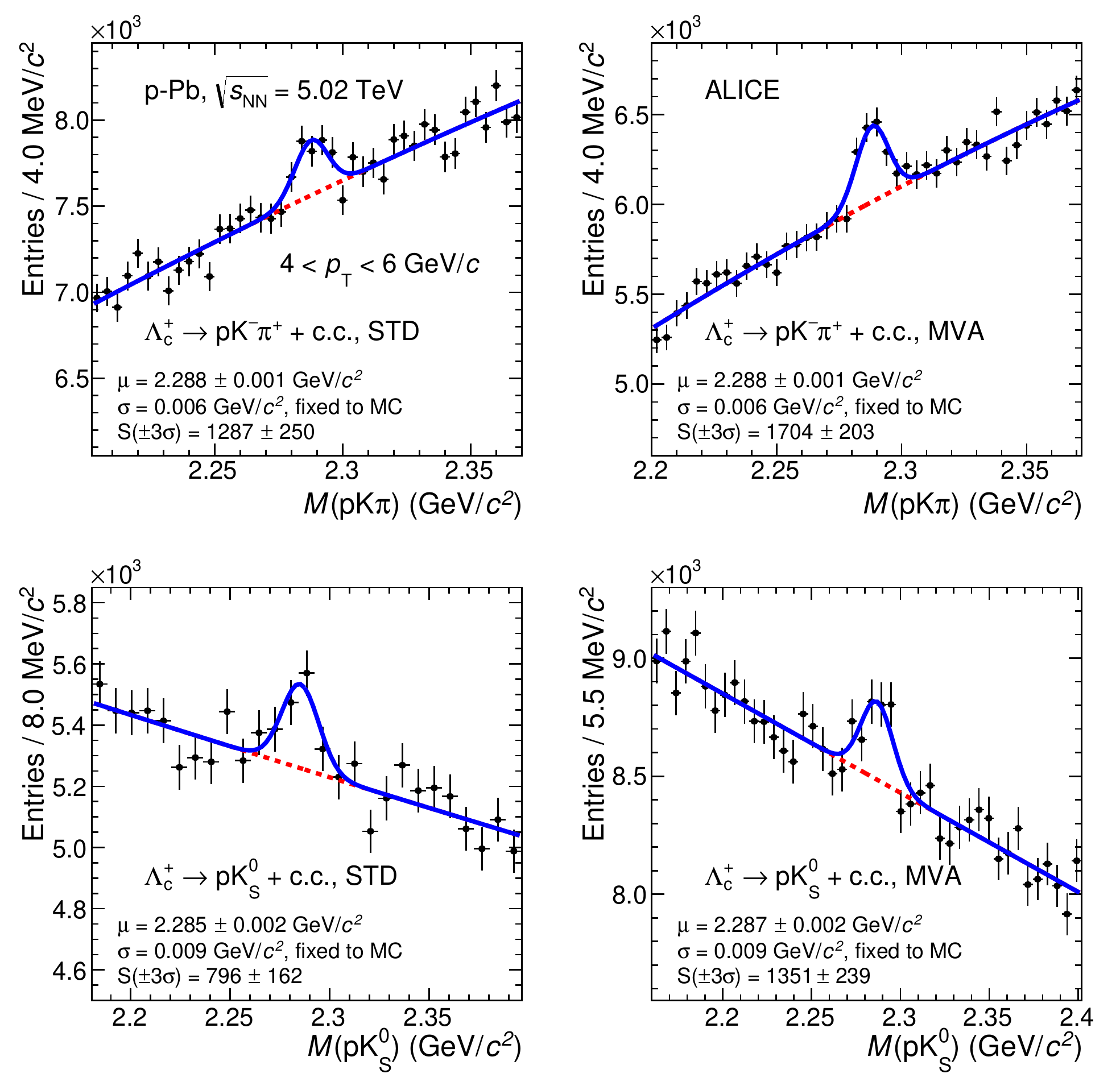}
\caption{Invariant-mass distribution of $\Lcplus$ candidates (and charge conjugates) for $4 <$ \pt $< 6$ \gevc in \pPb collisions at \sqrtsNN = 5.02~\tev. The dashed lines represent the fit to the background while the solid lines represent the total fit function. Top-left: $\LctopKpi$ STD analysis, top-right: $\LctopKpi$ MVA, bottom-left: $\LctopKzS$ STD analysis and bottom-right: $\LctopKzS$ MVA.}
\label{fig:InvMasspPb}
\end{figure}

Figures~\ref{fig:InvMasspp} and~\ref{fig:InvMasspPb} show examples of the invariant-mass distributions in one \pt interval for \pp and \pPb collisions, respectively for each of the methods discussed in this section.

\subsection{Semileptonic decay mode}
\label{sec:SemileptonicDecayMode}

The \Lc production cross section in \pp collisions at $\sqrt{s}=7$ TeV was also measured from its semileptonic decay \LctoenuLambda, based on reconstructed e$^+\Lambda$ pairs.
This analysis follows a procedure similar to the one presented in our recent work on the measurement of \xicz\ via its semileptonic decay, $\xicz\ \rightarrow$ e$^+\Xi^-\nu_{\rm e}$~\cite{Xic}.  
Here, we briefly describe the analysis approach for the \Lc with an emphasis on the differences from that analysis.

\Lcplus~candidates are defined from e$^+\Lambda$ pairs by combining a track originating from the primary vertex,  denoted electron track in the following, and a $\Lambda$ baryon reconstructed through the decay \Lambdatoppim, by exploiting the fact that its V$^0$-shaped decay topology is significantly displaced from the interaction point, given the additional lifetime of $\Lambda$ hyperons, $c\tau$ = 7.89 cm~\cite{Olive:2016xmv}. 
The V$^0$ candidates are built from pairs of tracks with $|\eta|<0.8$ reconstructed in the TPC and the ITS provided that they pass reconstruction quality criteria in a similar way as done for the hadronic decay channels. Additional cuts were applied to select the V$^0$-shaped decays: distance of closest approach between the daughter tracks smaller than 1 cm, $|d_0|$ of the daughter tracks larger than 0.06 cm, and cosine of the V$^0$ pointing angle to the primary vertex larger than 0.99. 
The compatibility of the p$\pi^-$ invariant mass with the  $\Lambda$-baryon mass within 8 MeV$/c^{2}$ was required in the analysis. 
The $\Lambda$ sample obtained with these selections is characterised by a signal-to-background ratio of about 20 for \pt $> 0$.  
Electron tracks were required to satisfy the reconstruction quality criteria described in \cite{Xic}.  
The PID selection was based, with respect to the electron hypothesis, on a $|n_\sigma|<3$ cut on the TOF signal and a \pt dependent $n_\sigma$ cut on the TPC signal:  $(-3.9+1.2\pt-0.094\pt^2)<n_\sigma<3$, with \pt expressed in \gevc.
The \pT-dependent lower limit for the TPC $n_\sigma$ is defined to have a constant purity over the measured \pt interval.  
Reconstructed 
e$^+\Lambda$ pairs were further required to have an opening angle smaller than 90 degrees and an invariant mass smaller than the \Lc mass.
 
\begin{figure}[t!]
\centering
\includegraphics[width=0.48\textwidth]{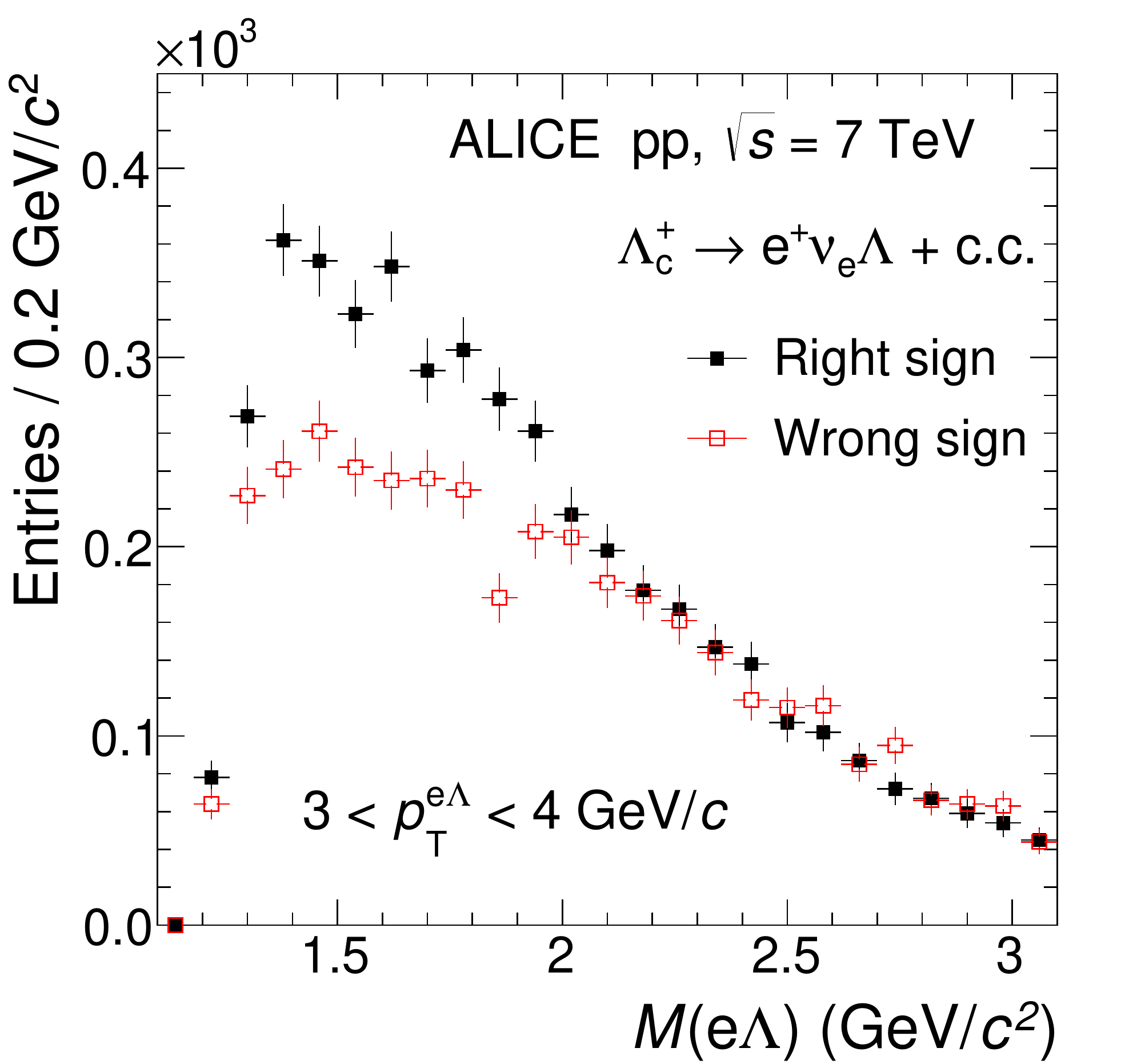}
\caption{Invariant mass distributions of e$\Lambda$ pairs for RS and WS combinations in the interval $3 < \pt^{{\rm e}\Lambda} < 4$ \gevc in pp collisions at $\sqrt{s}$~= 7 TeV.}
\label{fig:InvMassSemileptonic}
\end{figure}

Due to the missing neutrino, the invariant-mass distribution of e$\Lambda$ pairs does not show a peak at the \Lc mass and the raw yield cannot be extracted via a fit to the invariant-mass distribution with signal and background components as done for the hadronic
decay channels. 
Here, similarly to~\cite{Xic}, the background contributions were estimated using the fact that  \lambdac\ baryons decay only into e$^+\Lambda$ pairs, denoted as right-sign (RS), and not into e$^-\Lambda$ pairs, denoted as wrong-sign (WS), while background candidates contribute to both RS and WS pairs. 
The \Lc raw yield distribution was obtained by subtracting the WS contribution from the RS yields. 
Other contributions to e$\Lambda$ pairs, such as the contributions of \lambdab\ semileptonic decays to WS pairs and of \xiczp\ decays  to RS pairs, are corrected after the subtraction. The obtained \Lc raw yield in the intervals of e$\Lambda$-pair momentum  are further corrected for the missing momentum of the neutrino, as discussed below.  
Figure~\ref{fig:InvMassSemileptonic} shows the uncorrected e$\Lambda$ invariant-mass distributions for WS and RS pairs for the interval $3<\pt^{{\rm e}\Lambda}<4$~GeV/$c$. 

The \xiczp\ baryons contribute to RS  pairs through the decay chain $\xiczp\ \rightarrow$ e$^+\Xi^{-,0}\nu_{\rm e} \rightarrow$ e$^+\Lambda\pi^{-,0}\nu_{\rm e}$.
This contribution was estimated and subtracted from the RS yield to extract the yield of e$\Lambda$ pairs originating from \lambdac\ decays.
First, the ratio of e$\Lambda$ pairs from \xicz\ and \xicp\ was determined. 
Assuming that the production of \xicz\ and \xicp\ is the same, the difference in the e$\Lambda$ pair yields arises from their different branching ratios into the relevant decay modes.  
The ratio BR($\Xi^+_{\rm c}\rightarrow$ e$^+\Xi^0\nu_{\rm e}$)/BR($\Xi^0_{\rm c}\rightarrow$ e$^+\Xi^-\nu_{\rm e}$) 
 was measured by CLEO in e$^+$e$^-$ collisions below $\Upsilon$(4S) energies and found to be $ 2.46 \pm 0.7 ^{+0.33}_{-0.23}$~\cite{Alexander:1994hp}. 
Then, the relative contribution of \xiczp\ decays to the total yield of RS pairs was calculated. 
This was done using two different methods. 
In the first method, the \xiczp\ contribution in the $\pt^{\mathrm{e}\Lambda}$ distribution was calculated as

\begin{equation}
  N_i(\pt^{\mathrm{e}\Lambda}) = \sum_j F^{\Xi_{\rm c}^0}_{ij} M_j(\pt^{\Xi_{\rm c}^0}) + 2.46\cdot \sum_j F^{\Xi_{\rm c}^+}_{ij} M_j(\pt^{\Xi_{\rm c}^0}),
\label{eq:leyield}
\end{equation}

 where $N_i$ is the yield of e$\Lambda$ pairs in $i$-th $\pt^{\mathrm{e}\Lambda}$ bin, $M_j$ is the number of $\Xi_{\rm c}^{0}$ in $j$-th $\pt^{\Xi_{\rm c}^{0}}$ bin, which is computed from the measured $\Xi_{\rm c}^0$ cross section~\cite{Xic} as detailed below, and $F_{ij}^{\Xi_{\rm c}^{0,+}}$ are the matrices taking into account the reconstruction and selection efficiencies and the decay kinematics to convert $\pt^{\Xi_{\rm c}^{0,+}}$ into $\pt^{\mathrm{e}\Lambda}$. 
 
The \xicz\ cross section in the \pt\ range  $1< \pt < 8$ GeV/$c$ was taken from the measurement reported in~\cite{Xic} and the cross section outside the measured \pt\ range was estimated using the Tsallis function,

\begin{equation}
\frac{\mathrm{d}^2\sigma}{\mathrm{d}p_{\mathrm{T}}\mathrm{d}y} =  C\pt\Bigg[1+\frac{\sqrt{\pt^2+m^2}-m}{nT}\Bigg],
\label{eq:tsallis}
\end{equation}

where $C$ is a normalisation constant, $m$ is the $\Xi_{\rm c}^0$ baryon mass, and the parameters $n$ and $T$ were extracted from a fit to the data in the measured \pt range.  
The ratio between the yield of e$\Lambda$ pairs from $\Xi_{\rm c}$ decays and  that of inclusive e$\Lambda$ pairs was found to be independent of $\pt^{{\rm e}\Lambda}$ in the measured interval, with an average value of 0.38 $\pm$ 0.10, where the uncertainty also includes the contribution from the branching ratios measured by CLEO. 

The second approach exploits the fact that the distance between the interaction point and the decay vertex of  $\Lambda$ baryons originating from \Lc decays is on average smaller than that of $\Lambda$ baryons from $\Xi_{\rm c}$ decays, mediated by $\Xi$ hyperons ($c\tau\sim 4.91$ cm~\cite{Olive:2016xmv}).
In detail, for each $\pt^{\mathrm{e}\Lambda}$ interval, the \xiczp\ fraction was determined by fitting the measured distribution of the distance of the baryon decay point from the interaction point   
with the two contributions of $\Lambda$ baryons originating from \lambdac\ and \xiczp\ decays  
generated with {\sc pythia\small6.4.21} (Perugia-0 tune)~\cite{Skands:2009zm}.
Also in this case, no \pt\ dependence of the \xiczp\ relative contribution in the yield of e$\Lambda$ pairs was observed, and the average was found to be 0.52 $\pm$ 0.09, consistent with the result from the first approach.
By taking the weighted average of the values obtained with the two methods, we obtained 0.46 $\pm$ 0.06 as the relative contribution of \xiczp\ decays. 

 \lambdab\ baryons contribute to WS pairs through their decay mode $\lambdab\ \rightarrow$ e$^-\lambdac\bar{\nu}_{\rm e} $, with BR (10.3 $\pm$ 2.2)\%, followed by the subsequent decay $\lambdac\rightarrow \Lambda + X$, with BR (35 $\pm$ 11)\%~\cite{Olive:2016xmv}. 
This contribution was estimated using the \lambdab\ measurement at central rapidity by CMS~\cite{Chatrchyan:2012xg}, which covers the transverse momentum interval \pt $>10$ GeV/$c$.  
The cross section for $\pt <  10$ GeV/$c$ was estimated  using the Tsallis parameterisation reported in~\cite{Chatrchyan:2012xg}. 
The \lambdab\ distribution was further converted into an e$\Lambda$ distribution via simulations, taking into account the detector acceptance, the reconstruction and selection efficiency, and the decay kinematics to determine the fraction of \lambdab\ momentum carried by e$\Lambda$ pairs. 
The obtained yield of e$\Lambda$ pairs originating from \lambdab\ decays was added to the measured e$\Lambda$ yield after the WS pairs were subtracted.
The correction is found to increase with $\pt^{{\rm e}\Lambda}$ reaching about 10\% in the highest $\pt^{{\rm e}\Lambda}$ interval.

The correction for the missing momentum of the neutrino was performed 
by using the response matrix determined with the full detector simulation of {\sc pythia} events containing \Lc baryons and using the Bayesian unfolding technique~\cite{D'Agostini:1994zf} implemented in the {\sc RooUnfold} package~\cite{Adye:2011gm}. 
The number of iterations, which is a regularisation parameter of the Bayes unfolding, was chosen to be 3 in this analysis. It was verified that the final result is not sensitive to this choice.

\section{Corrections}
\label{sec: Corrections}

The \pt-differential cross section of prompt \Lcplus ~baryon production was obtained for each decay channel as:

\begin{equation} 
\label{eqn:lc_normalization}
\frac{\mathrm{d^2}\sigma^{\Lambda_{\rm c}^+}}{\mathrm{d}\pt\mathrm{d}y} = \frac{1}{2 c_{\Delta y} \Delta \pt} \frac{1}{\rm{BR}}
\frac{  f_{\rm prompt} \cdot N^{\Lambda_{\rm c}}_{\lvert y \lvert < y_{\rm fid}}}{ (A\times\varepsilon)_{\rm prompt}} \frac{1}{\lumi},
\end{equation}

where $N^{\Lambda_{\rm c}}$ is the raw yield (sum of particles and antiparticles) in a given $\pt$ interval with width $\Delta \pt$, $f_{\rm prompt}$ is the fraction of the raw yield from prompt \Lc, 
$(A\times\varepsilon)$ is the product of acceptance and efficiency for prompt \Lc baryons, BR is the branching ratio for the considered decay mode and $\lumi$ is the integrated luminosity.
The correction factor for the rapidity coverage $c_{\Delta y}$ was computed, for the hadronic decay modes,
as the ratio between the generated \Lc-baryon yield in 
$|y_{\rm lab}|<y_{\rm fid}$(\pt) and that in $|y_{\rm lab}|<0.5$.
For the semileptonic decay analysis, the rapidity of the \Lc candidate cannot be calculated due to the missing neutrino momentum, 
and the $y_{\rm fid}$ cut cannot be applied. A factor $c_{\Delta y}$ = 1.6 was used in this case assuming a flat distribution of the 
\Lc candidates in $|y_{\rm lab}|<0.8$, which was verified with an accuracy of  1\% using pure Monte Carlo information from {\sc pythia}.
The factor 2 in the denominator of Eq.~\ref{eqn:lc_normalization} takes 
into account that the raw yield is the sum of particles and antiparticles, while the cross section is given for particles only and is computed as the 
average of \Lcplus~and \Lcminus.

The correction for the detector acceptance and reconstruction efficiency $(A \times\varepsilon)$ was 
obtained following the same approach as discussed in~\cite{ALICE:2011aa}. 
The correction factors were obtained from Monte Carlo simulations where the detector
and data taking conditions of the corresponding data samples were reproduced.

Contrary to the case of \pp collisions, for which the simulation describes in a satisfactory way the charged-particle multiplicity in data,
in \pPb collisions a weighting procedure based on the
event multiplicity was applied in the calculation of the efficiency from the si\-mulated events. This approach accounts for the dependence of the 
efficiency on the event multiplicity, which is due to
the fact that the resolutions of the primary 
vertex position and of the variables used in the geometrical selections of displaced decay vertices improve with increasing multiplicity.

The efficiency was computed separately for prompt and non-prompt \Lc (originating from \Lb-baryon decays).
The \LctopKpi decay channel includes not only the direct (non-resonant) decay mode, but also three resonant channels, namely
$\mathrm{p\overline{K}}^*(892)^0$, $\Delta(1232)^{++}\mathrm{K}^-$ and $\Lambda(1520)\pi^+$. 
The kinematical properties of these decays are different, resulting in different
acceptances and efficiencies for each case. The final correction was determined
as a weighted average of the $(A\times\varepsilon)$ values of the four decay channels, using the relative branching ratios as weights.

\begin{figure}[t!]
\centering
\includegraphics[width=1.0\textwidth]{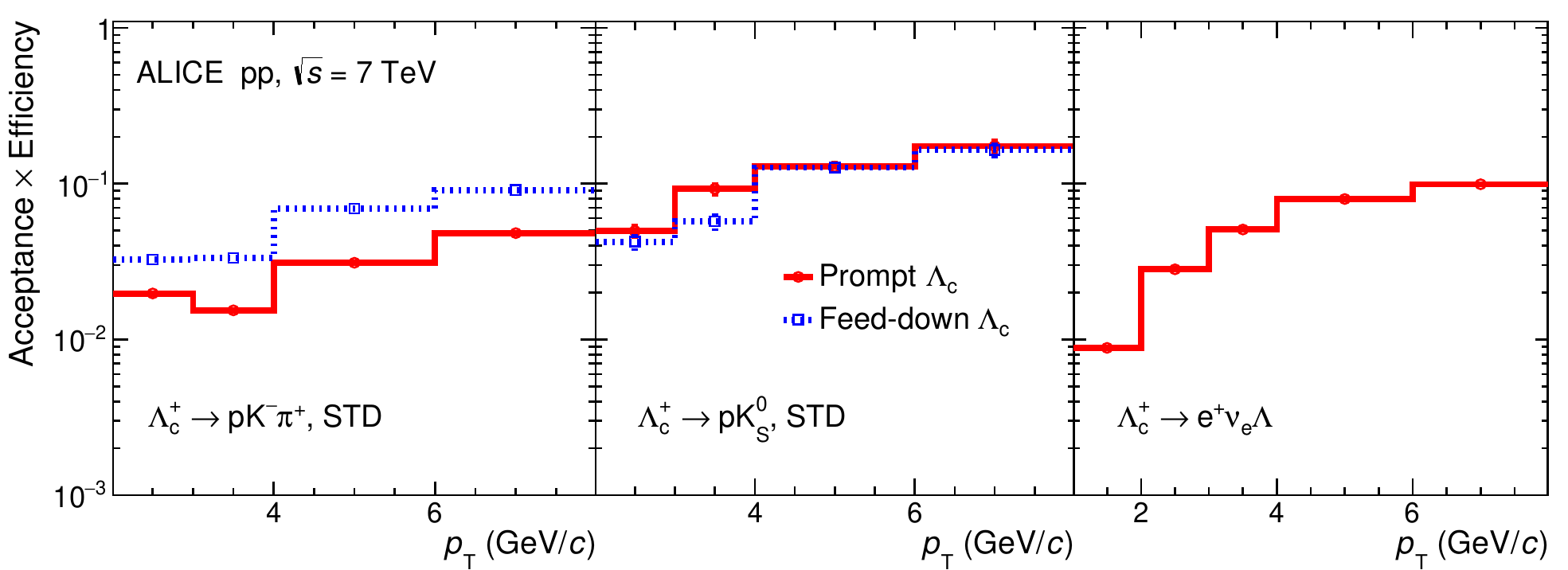}
\caption{Product of acceptance and efficiency 
for \Lc in pp collisions at $\sqrt{s}=7~\TeV$,
  as a function of \pt. From left to right: \LctopKpi, \LctopKzS, and \LctoenuLambda. For hadronic decays the solid lines correspond
  to the prompt \Lc, while the dotted lines represent $(A \times \epsilon)$ for \Lc baryons originating from beauty-hadron decays. The efficiency for semi-leptonic decays (same for both prompt and non-prompt \Lc) is represented with one solid line. The statistical uncertainties are smaller than the marker size. }
\label{fig:AccEff_pp7TeV}
\end{figure}

Figure~\ref{fig:AccEff_pp7TeV} shows the product of acceptance times efficiency $(A\times\varepsilon)$ for \Lc
baryons with $|y| < y_{\rm fid}$(\pt)
in pp collisions 
at $\sqrt{s}=7~\TeV$, as a function of transverse momentum, for \LctopKpi (left panel), \LctopKzS (middle panel), and \LctoenuLambda (right panel). The higher 
efficiency for \Lc from beauty-hadron decays in the \LctopKpi decay channel 
is due to the geometrical selections on the displaced decay-vertex topology, which enhance the non-prompt component 
because of the additional lifetime of the beauty hadrons.
In the case of the \LctopKzS decay,
for \pt $ <4 $ \gevc the efficiency for prompt \Lc is slightly higher
because the upper cut applied on the bachelor $d_{0}$ to remove secondary tracks
rejects preferentially \Lc from beauty-hadron decays.
In the semileptonic analysis no selection is made on variables related to the displacement of the $\Lc$ decay vertex from the primary vertex, and therefore
the efficiency is the same for both prompt and non-prompt $\Lc$.

\begin{figure}[t!]
\centering
\includegraphics[width=0.8\textwidth]{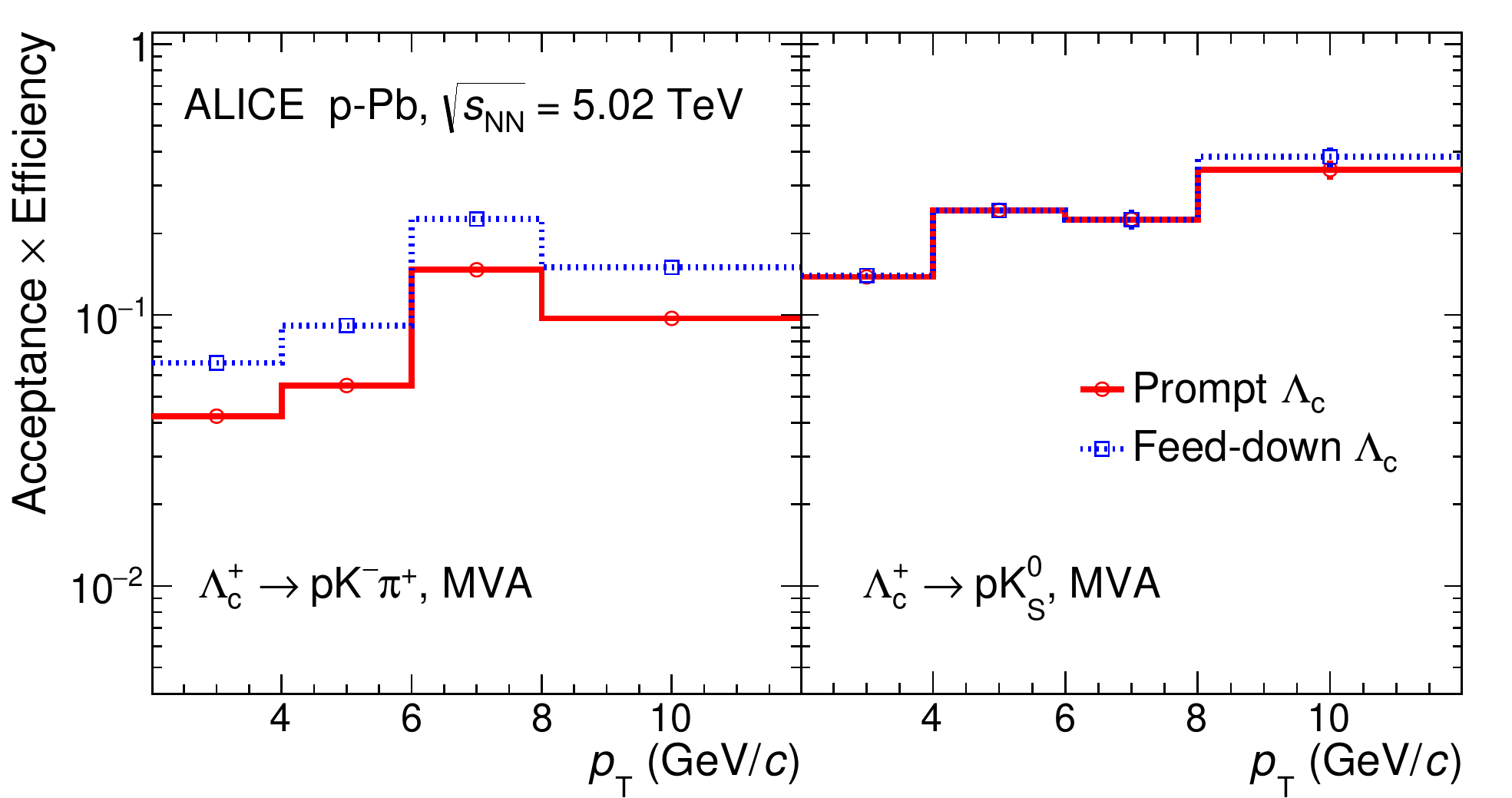}
\caption{Product of acceptance and efficiency
  for the two \Lc hadronic decay channels in p-Pb collisions at \sqrtsNN = 5.02 \tev, as a function of \pt with the MVA technique. From left to right: \LctopKpi and \LctopKzS. The solid lines correspond
  to the prompt \Lc, while the dotted lines represent $(A \times \epsilon)$ for the \Lc from beauty-hadron decays. The 
  statistical uncertainties are smaller than the marker size.}
\label{fig:AccEff_pPb5TeV}
\end{figure}

When using the Multivariate Analysis approach, a further correction factor ($\epsilon_{\rm BDT}$) was required. This additional ingredient corresponds to the BDT cut efficiency, quantifying the fraction of true \Lc candidates accepted by the selection on the classifier output. Since the BDT analysis employed a different set of pre-selections, a specific correction factor $\epsilon_{\rm presel}$ for those was also taken into account. The final efficiency correction is $\epsilon=\epsilon_{\rm BDT}\times\epsilon_{\rm presel}$. 

The BDT cut efficiency was determined from the simulations with the {\sc pythia} and {\sc hijing} event generators described above by applying the classification algorithm resulting from the training of the BDT on the simulated sample enriched with \Lc described in Sec.~\ref{sec:HadronicDecayModes}.

The efficiency and acceptance corrections
for prompt and non-prompt \Lc in \pPb collisions are reported in Fig.~\ref{fig:AccEff_pPb5TeV} 
as a function of \pt in the rapidity range 
$|y_{\rm lab}| < y_{\rm fid}(\pT)$
for the decay channels \LctopKpi (left panel) and \LctopKzS (right panel) for the MVA technique.
The non-monotonic trend seen in the efficiencies for both channels is a result of the non-monotonic tightness of the BDT cut chosen as a function of \pt, and it was verified that these choices do not have a significant systematic effect on the result.

To obtain the factor $f_{\textrm{prompt}}$, i.e. the fraction of prompt \Lc in the raw yield,
the production cross section of \Lc from \Lb decays
was estimated using the beauty hadron \pt shape from FONLL~\cite{Cacciari:1998it,Cacciari:2012ny} as described in detail in~\cite{ALICE:2011aa}
(the contribution from B-meson decays to \Lc was checked and found negligible \cite{Olive:2016xmv}). 
The fraction of beauty quarks that fragment to beauty hadrons and subsequently decay into \Lc baryons
$f(\rm{b} \rightarrow \Lc)$ = 0.073 was taken from \cite{Gladilin:2014tba} and the \Lb $\rightarrow$ \Lc + {\it{X}} decay kinematics were modelled using the {\sc EvtGen} \cite{Lange:2001uf} 
package. 
The production cross section of \Lc from \Lb was then multiplied for each decay channel in each \pt interval by $(A \times \epsilon)_{\textrm{feed-down}}$, 
the factor $c_{\Delta y}$, the branching 
ratio BR and the integrated luminosity $\lumi$.
The correction factor $f_{\textrm{prompt}}$ was calculated in pp collisions as:

\begin{equation}	
f_{\textrm{prompt}} = 1 - \frac{N^{\Lambda_c\rm{feed-down}}}{N^{\Lambda_c}} = 1 - \frac{(A \times \epsilon)_{\rm feed-down} ~c_{\Delta y}~\Delta p_{\textrm{T}} ~{\rm{BR}} ~{\lumi}}{N^{\Lambda_c}/2} \times \Big(\frac{\textrm{d}^{2}\sigma}{\textrm{d}p_{\textrm{T}}\textrm{d}y}\Big)^{\rm FONLL}_{\rm feed-down}.
\label{eq:nb}
\end{equation}

where $N^{\Lambda_c}/2$ is the raw yield, which was divided by a factor of two to account for particles and antiparticles.

For \pPb collisions, a hypothesis on the nuclear modification factor $R_{\textrm{pPb}}^{\textrm{feed-down}}$ of \Lc from beauty--hadron decays
was added as an additional factor in the last term of Eq.~\ref{eq:nb}. As in the D-meson analyses \cite{Adam:2016ich}, it was assumed that the $\RpPb$ of prompt and feed-down \Lc were equal and their ratio was varied in the range $ 0.9 < R_{\textrm{pPb}}^{\textrm{feed-down}}/R_{\textrm{pPb}}^{\textrm{prompt}} < 1.3$ to evaluate the systematic uncertainties.
The values of $f_{\textrm{prompt}}$ range between 95\% 
and 99\% depending on the decay channel and \pt.

\section{Evaluation of systematic uncertainties}
\label{sec: Systematics}
This section is dedicated to the description of the various sources of systematic uncertainties
for each analysis presented here. 
First, the systematic uncertainties for the \Lc hadronic decay modes in both \pp and \pPb collisions will be discussed.
Then, the systematic uncertainties studied for the \Lc semileptonic decay mode will be presented. 
For each analysis, the different sources of systematic uncertainties were assumed to be uncorrelated among each other
and the total systematic uncertainty was determined in each \pt interval as the quadratic sum of the different contributions.

A summary of the systematic uncertainties is shown in \cref{Syspp,SyspPb,systlc}, 
for the hadronic analyses in \pp collisions, the hadronic analyses in \pPb collisions, and the semileptonic analysis in \pp collisions, respectively.
These include the uncertainties specific to each analysis
as well as the uncertainties associated to the branching ratios of the \Lc decay modes~\cite{Olive:2016xmv}.
The measured cross sections are also affected by a global normalisation uncertainty related to the determination
of the integrated luminosity of 3.5\%~\cite{Abelev:2012sea} and 3.7\%~\cite{Abelev:2014epa} in \pp and \pPb collisions, respectively. 

\subsection{Systematic uncertainties for the hadronic channels}
\label{sec: Systematics uncertainties for the hadronic channels}
The systematic uncertainty on the raw-yield extraction was estimated for each decay mode and in each \pt interval by repeating the fit to the invariant-mass distributions under different approaches. The following variations to the fit procedure were considered: (i) the background function, for which three different functions were tested (parabolic, linear and exponential), and 
(ii) the lower and upper limit of the fit range of the invariant-mass distributions.
For each combination of the
aforementioned variations, the fit was performed under different assumptions on the width and position 
of the Gaussian function modelling the \Lc peak in the invariant-mass distributions, namely: (a) fixing the Gaussian width to the value obtained from simulation (used as default); (b) fixing the peak position to the value obtained from simulations; (c) leaving the peak width and position as free parameters of the fit; (d) fixing both the peak width and position. 
Only those cases satisfying quality criteria on the resulting fits 
were considered to assess the final systematic uncertainty, which was defined as the RMS of the distribution of the signal yields obtained from the different 
trials.

The contribution to the systematic uncertainty due to the tracking efficiency was evaluated as discussed in~\cite{Acharya:2017jgo} for the D-meson analysis, i.e. by comparing the probability of matching TPC tracks to ITS points in data and simulation and by varying the quality cuts to select the tracks used in the analysis. The uncertainty on the matching efficiency was defined as the relative difference of the matching efficiencies in data and simulations.
The matching efficiency for primary tracks is higher than that for secondary tracks produced far from the interaction point in strange particle decays (such as those coming from the \Kzs decay in the \LctopKzS channel) or in interactions with the detector material.
Different fractions of primary and secondary tracks, in data and simulations, could lead to a wrong estimation of the systematic uncertainty for the matching. 
For this reason, the comparison of the matching efficiency in data and simulations was done after weighting the relative abundances of particles in Monte Carlo to match those observed in data.
The uncertainty resulting from these studies was added in quadrature with the uncertainty on the track selection for the final uncertainty on the tracking efficiency.

Systematic uncertainties on the efficiency can also arise from possible differences in the distributions and resolution of selection variables between data and the simulation.
The systematic effect induced by these imperfections was estimated 
by repeating the analysis with several sets of selection criteria for the \Lc candidates. Each selection was varied 
with respect to the central value, obtaining a relative variation of the efficiency between 5\% and 40\%.
The uncertainty due to these selections was then estimated from the RMS of the cross sections resulting from all the variations
and it ranges from 4\% to 10\% depending on the analysis and the decay channel. 

The results presented in this paper rely on an extensive use of the PID capabilities of the TPC and TOF detectors. The uncertainties
arising from discrepancies in the PID efficiency in data and simulation were estimated by varying the PID strategy (with tighter or looser $n_\sigma$ cuts, or with different configurations for the Bayesian PID, for the \LctopKzS and \LctopKpi analyses, respectively),
and estimating the uncertainty from the RMS of the resulting corrected yields obtained from the tests. 

The efficiencies determined from the simulations depend on the generated \pt distribution of \Lc baryons.
The central values of the correction factors were obtained by re-weighting the \Lc distribution generated by {\sc pythia} according to the ratio of the \pt distribution of D$^0$ mesons from FONLL calculations and from {\sc pythia} simulations.
A systematic uncertainty  was defined by considering the RMS variation of the efficiencies determined with different generated 
\pt shapes, namely: (i) c-quark \pt distributions from FONLL, (ii) \Lc \pt shapes from {\sc pythia}.
It was found to be 3\% at most, depending on the analysis. 

As discussed in Sec.~\ref{sec: Corrections}, the efficiency for \Lc reconstruction and selection depends on the multiplicity of particles produced in the collision, since the resolution on the primary vertex improves with increasing multiplicity. 
For p--Pb collisions, a systematic uncertainty was assigned to account for the accuracy of the multiplicity weighting procedure applied in the efficiency calculation. It amounts to 1\% for the analysis using MVA, while it is negligible for the STD analysis, for which the efficiency shows a less pronounced dependence on multiplicity.

The contribution to the uncertainties coming from the subtraction of \Lc baryons from \Lb decays was calculated as the envelope of the uncertainty bands obtained (i) by varying the $\pt$-differential cross section of beauty hadrons within the theoretical uncertainties of the FONLL calculation, and (ii) with the same
method but after scaling by a factor of two the fraction $f({\rm b} \rightarrow \Lc)$, which is used together with FONLL cross sections to determine the yield of \Lc from \Lb decays.
The uncertainty in the FONLL calculation of (i) was determined by changing the b-quark mass and the perturbative scales, 
as explained in \cite{Cacciari:2012ny}, also including the uncertainty on $f({\rm c} \rightarrow \Lb)$ from~\cite{Gladilin:2014tba},
and finally adding in quadrature the uncertainty estimated for the used PDF set. 
The variation by a factor of two of the fraction $f({\rm b} \rightarrow \Lc)$ in (ii) was motivated by the observation that FONLL calculations describe
the available \Lb cross section measurements in \pp collisions at \sqrts=7 TeV once the value of 0.197 measured at CDF~\cite{Aaltonen:2008zd} is taken. As noted in
Sec.~\ref{sec: Introduction} the different values of this fragmentation
fraction measured in hadron--hadron collisions with respect to \ee interactions has been interpreted as a violation of its universality~\cite{Olive:2016xmv}. If the value $f({\rm b} \rightarrow \Lb)$ = 0.088, derived from LEP measurements in electron-positron collisions~\cite{Gladilin:2014tba}, is used for the fragmentation fraction, the FONLL calculations underestimate by a factor of about two 
the \Lb measurement by LHCb at forward rapidity in the same \pT region of this analysis~\cite{Aaij:2015fea}
and by a factor of about 1.6 the CMS measurements at mid-rapidity in their
lowest reported \pT interval ($10<\pt<13$ ~\gevc)~\cite{Chatrchyan:2012xg},
corresponding to the high-\pt region of this analysis.

Additional possible sources of systematic uncertainties were checked. The difference between the 
resolution on the \Kzs mass in data and simulation, the difference in the reconstruction efficiencies for \Lcplus and \Lcminus,
and the possible contamination in the \Lc invariant-mass distribution coming from
\DtopiKzs and \DstoKKzs decays were all checked and 
proved to give a negligible contribution to the final uncertainties. These decays enter the candidate \Lc sample only if the kaon or the pion passes the proton PID selection.

\begin{table}[!htb]
  \begin{center}
    \def\arraystretch{1.2}\tabcolsep=2pt    
    \begin{tabular}{lccccc} 
    \toprule
       & \multicolumn{2}{c}{$\LctopKpi$} & & \multicolumn{2}{c}{$\LctopKzS$} \\
       \cmidrule{2-3} \cmidrule{5-6} 
       & lowest \pt & highest \pt & & lowest \pt & highest \pt \\
       \midrule
       Yield extraction (\%) & 11 & 4 &  & 7 & 9 \\
       Tracking efficiency (\%) & 4 & 3 & & 7 & 5 \\
       Cut efficiency (\%) & 11 & 12 & & 5 & 6 \\
       PID efficiency (\%) & 4 & 4 & & 5 & 5 \\
       MC \pt shape (\%) &  2 & 2  & & neg. & 1.5 \\
       Beauty feed-down (\%)  & $\substack{+1 \\ -4}$ &  $\substack{+2 \\ -11}$ & & $\substack{\mathrm{neg.} \\ -2}$ & $\substack{+1 \\ -4}$ \\       
        Branching ratio (\%) & \multicolumn{2}{c}{5.1} & & \multicolumn{2}{c}{5.0} \\
        Luminosity (\%) & \multicolumn{5}{c}{3.7} \\
      \midrule
    \end{tabular}
  \end{center}
  \caption{Summary of relative systematic uncertainties for the lowest and highest \pt intervals considered in the analysis, for the two \Lc hadronic decay modes in \pp collisions. When the uncertainty was found to be $< 1\%$, it was considered negligible (``neg.'' in the table).}
    \label{Syspp}
\end{table} 

\begin{table}[!htb]
  \begin{center}
    \def\arraystretch{1.2}\tabcolsep=2pt  
    \begin{tabular}{lccccccccccc} 
    \toprule
       & \multicolumn{5}{c}{$\LctopKpi$} & & \multicolumn{5}{c}{$\LctopKzS$} \\
       \cmidrule{2-6} \cmidrule{8-12} 
       & \multicolumn{2}{c}{STD} & & \multicolumn{2}{c}{MVA} & & \multicolumn{2}{c}{STD} & & \multicolumn{2}{c}{MVA} \\
       \cmidrule{2-3} \cmidrule{5-6} \cmidrule{8-9} \cmidrule{11-12}  
       & lowest & highest & & lowest & highest & & lowest & highest & & lowest & highest\\
       & \pt & \pt & & \pt & \pt & & \pt & \pt & & \pt & \pt \\  
       \midrule
        Yield extraction (\%) & 10 & 11 & & 7 & 4 & & 10 & 10 & & 11 & 8 \\
        Tracking efficiency (\%) & 10 & 7 & & 10 & 7 & & 10 & 6 & & 10 & 6 \\
        Cut efficiency (\%) & 9 & 12 & & 8 & 6 & & 5 & 7 & & 5 & 8 \\
        PID efficiency (\%) & 6 & 6 & & neg. & neg. &  & 6 & 6 &  & neg. & neg. \\
        MC \pt shape (\%) & 2 & 2 &  & neg. & 3  & & 1 & 3 & & neg.  & neg. \\
        Multiplicity (\%) & neg. & neg. & & neg. & neg. &  & neg.  & neg. & & 1 & 1 \\
        Beauty feed-down (\%) & $\substack{+1 \\ -5}$ & $\substack{+2 \\ -10}$ & &  $\substack{+1 \\ -5}$ & $\substack{+2 \\ -10}$ & & $\substack{\mathrm{neg.} \\
          -3.}$ & $\substack{+2 \\ -7}$ & & $\substack{\mathrm{neg.} \\ -3}$ & $\substack{+2 \\ -7}$ \\
        Branching ratio (\%) & \multicolumn{5}{c}{5.1}  & \multicolumn{5}{c}{5.0}\\
        Luminosity (\%) &  \multicolumn{11}{c}{3.5} \\
      \midrule
    \end{tabular}
  \end{center}
  \caption{Summary of relative systematic uncertainties for the lowest and highest \pt intervals considered in the analysis 
    for the two \Lc hadronic decay modes and the two analysis techniques in \pPb collisions. When the uncertainty was found to be $< 1\%$, it was considered negligible (``neg.'' in the table).}
    \label{SyspPb}
\end{table}

For the analyses using MVA, specific sources of systematic uncertainty were additionally considered. The uncertainty associated to the selection on the MVA classifier output was estimated by repeating the analysis with different cutting points after verifying that these variations induce a significant modification of the efficiency, between 10 and 40\%. 
The RMS of the distribution of the corrected yields was then used to assign the systematic uncertainty (reported under cut efficiency in Tab.~\ref{SyspPb}).

A possible systematic effect of the specific multivariate algorithm chosen (BDT)~\cite{Hocker:2007ht} was checked by changing the configuration of the MVA method. 
These changes included the number of trees used to construct the forest, the maximum depth of the trees constructed, the boosting algorithm, the application of data preprocessing
such as the transformation of input variables to reduce correlation or the transformation of the variable shapes into more appropriate forms, 
the metric defining the separation criterion in the node and the number of input variables. The effects of such modifications in the corrected yields were found to be negligible.

The PID-related variables play an important role in the multivariate selection, since they offer the largest discrimination power.
As a further cross-check, the systematic uncertainty associated with the inclusion of these variables in the multivariate 
selection was estimated. For the \LctopKpi analysis, the kaon and pion priors used in the calculation of the Bayesian probability were modified conservatively based on the maximum mismatch
between the default priors determined through an iterative procedure and the measured particle abundances \cite{Adam:2016acv}, and the BDT efficiency was determined for each modification. For the \LctopKzS analysis, the Bayesian probability for protons in simulation and data 
was compared using the daughter particles of V0 decays
in order to select a pure proton sample. The resulting variations were found to be 
consistent within 2--4\%; this effect was not included as an additional uncertainty source, since it should be accounted for in the 
BDT cut variation and its magnitude is smaller than the assigned systematic uncertainty. 
Moreover, to assess whether the Bayesian approach used in the MVA might lead to biased results, 
the \LctopKzS analysis was repeated using an $n_{\sigma}$ approach for the bachelor PID and not considering any PID in the BDT.
The results for the three cases were found to be compatible, and therefore no systematic uncertainty was assigned. As reported in~\secref{sec:HadronicDecayModes} and~\secref{sec: Corrections}, a loose particle 
identification, based on rectangular $n_{\sigma}$-compatibility cuts on the TPC and TOF PID response for pion, kaon and proton tracks 
is applied prior to the BDT. The systematic uncertainty associated with this cut was studied 
by comparing the corrected \Lc yield obtained with and without this cut (\LctopKzS) 
and without the TOF selection (\LctopKpi) and was found to be negligible in the \pt range considered here.

The contribution of the uncertainty related to the imperfect description of the impact parameter resolution in the simulation, which could affect the input variables related to vertex reconstruction, was checked
in the \LctopKpi analysis. For this check, the distribution of the input variables was altered by smearing the reconstructed track parameters to match the impact parameter resolution observed in data, and the BDT cut efficiency was recalculated. The change in efficiency was 2\% at low \pt, and less than 1\% at high \pt, consistent with being a contribution to the systematic uncertainty estimated with the cut-variation procedure.

\subsection{Systematic uncertainties for the semileptonic channel}
\label{sec: Systematic uncertainties for the semileptonic channel}

The following contributions to the systematic uncertainty on the \Lc cross section measurements through the $\lambdac\rightarrow {\rm e}\Lambda\nu_{\rm e}$ decay channel were considered: raw-yield extraction, $(A\times\varepsilon)$ correction factor, correction for the missing neutrino momentum and for  feed-down from beauty-hadron decays. These contributions were added in quadrature to obtain the total systematic uncertainty in each \pt\ interval and they are summarised
in Tab.~\ref{systlc}.

The systematic uncertainty due to the raw-yield extraction includes the uncertainties in the WS subtraction procedure, the estimation of the \xiczp\ contribution to RS pairs and the \lambdab\ contribution in WS pairs. 
The WS pair subtraction described in Sec. \ref{sec:SemileptonicDecayMode} was based on the assumption  that there were no charge asymmetric background sources and that the acceptance of RS and WS pairs were the same. 
The influence of the charge asymmetric background sources was evaluated using {\sc pythia} events with full detector simulation, as done in the \xicz\ analysis~\cite{Xic}, and found to be about 2\%. 
The difference in the acceptance of RS and WS pairs was estimated using a mixed-event technique and found to be negligible for this analysis. 
In addition, the impact on the background subtraction of the hadron contamination in the electron sample and the signal-to-background 
ratio was studied varying the electron identification criteria.  
The corrected spectra were all found to be consistent with the one obtained with the default selections and no systematic uncertainty was assigned. 

The \xiczp\ contribution to the RS pairs calculated as described in Sec.~\ref{sec:SemileptonicDecayMode} also contributes to the 
systematic uncertainty on the raw-yield extraction.  
An additional uncertainty of 10\% estimated from {\sc pythia} simulations
was assigned to take into account the possible \pt\ dependence of the fraction of \lambdac\ from \xicz\ decays 
and summed in quadrature to the value reported in Sec. \ref{sec:SemileptonicDecayMode}. 
The systematic uncertainty on the \lambdab\ contribution in WS pairs was estimated by taking into account the uncertainty on the \lambdab\ cross section measured by CMS~\cite{Chatrchyan:2012xg} and the uncertainty on the relevant branching ratios. 
The uncertainty was found to increase with \pt\ reaching about 5\% in the highest \pt\ interval. 

The $(A\times\varepsilon)$ factor could be affected by imperfections in the description of the detector alignment and response in the simulation.   
The systematic uncertainties due to the reconstruction and selection efficiency were estimated by repeating the analysis with different selection criteria for electrons, $\Lambda$ baryons, e$\Lambda$ pairs and by comparing the corrected yields. 
The systematic uncertainty on the electron reconstruction and selection efficiency was estimated via variations of the track-quality criteria and the PID selections for electron identification. 
The RMS of the \Lc corrected yields, which amounted to 4\% (track-quality) and 3\% (PID), was assigned as a systematic uncertainty. 
Similarly, a systematic uncertainty of 1\% on the $\Lambda$ reconstruction and selection efficiency, was estimated from the RMS of the inclusive $\Lambda$ corrected yield against variations of the criteria applied to select the $\Lambda$ decay tracks and its V$^0$ decay topology. 
In addition, a systematic uncertainty of 4\% on the $\Lambda$ efficiency due to possible imperfections in the description of the detector material in the simulations was considered and summed in quadrature to the one estimated from the variation of the selection criteria.  
The uncertainties on the electron and $\Lambda$ reconstruction efficiency were considered as correlated and combined linearly. 
The uncertainty on the e$\Lambda$ pair selection efficiency was estimated by varying the selection criteria on the opening angle and the invariant mass of the pair and a systematic uncertainty ranging from 1 to 25\% was assigned depending on \pt. 
The systematic uncertainty due to a possible imperfect description of the acceptance of e$\Lambda$ pairs in the simulation was estimated to be 11\% by comparing the azimuthal distribution of inclusive electrons and $\Lambda$ baryons in data and in the simulation. 
The uncertainty on the e$\Lambda$ pair acceptance was summed in quadrature to the one on the electron, $\Lambda$ and e$\Lambda$-pair selection efficiencies.

\begin{table}[t!]
\begin{center}
\def\arraystretch{1.2}\tabcolsep=6pt  
\begin{tabular}{lcc}	
\toprule
  & lowest \pt & highest \pt \\
\midrule
Raw-yield extraction (\%) & 17 & 17 \\
$(A\times\varepsilon)$ (\%) & 28 & 13 \\
Missing neutrino momentum (\%) & 3 & 11 \\
Beauty feed-down (\%) &  ${}^{\mathrm{neg.}}_{\mathrm{neg.}}$ & ${}^{+1}_{-7}$ \\ 
Branching ratio (\%) & \multicolumn{2}{c}{11}\\
Luminosity (\%) &  \multicolumn{2}{c}{3.7} \\
\midrule
\end{tabular}
\end{center}
\caption{Summary of relative systematic uncertainties for the \LctoenuLambda analysis in \pp collisions. The uncertainties smaller than 1\% are considered negligible (``neg.'' in the table).}
\label{systlc}
\end{table}

The dependence of the corrected results on the unfolding procedure was tested by (i) using as prior for the Bayesian unfolding the \pt\ distribution from {\sc pythia}
Monte Carlo simulations, and (ii) adopting different unfolding methods ($\chi^2$ minimisation with regularisation \cite{Grosse-Oetringhaus:2011tgu,unfolding_chi2} and Singular Value Decomposition \cite{Hocker:1995kb}). 
The RMS of the corrected yields was used to estimate the resulting uncertainty, which increases from 3\% to 11\% towards higher \pt. 

The uncertainty arising from the subtraction of the feed-down from  beauty-hadron decays was calculated in the same way as for the hadronic decays. 

\section{Results}
\label{sec: Results}

In this section, results are first presented in~\secref{sec:xsecResult} for the prompt \Lcplus~production cross sections in \pp and \pPb collisions obtained using the procedure discussed in Secs.~\ref{sec: Analysis overview and methods}-\ref{sec: Systematics}.
In the decay modes under study in \pp collisions, it was possible to extract a stable signal in the lowest \pT interval ($1< \pT < 2$ \gevc) only via the semileptonic decay. In the highest \pT interval ($6 < \pT < 8$ \gevc) it was not possible to extract a signal from the \LctopKpi invariant mass distribution. For \pPb collisions in the two hadronic decay modes under study with two different analysis methods (standard cuts and MVA) it was possible to extract a signal in four \pT intervals from 2 to 12 \gevc.

The results from each decay mode and analysis method agree within statistical and systematic uncertainties. After averaging the results obtained from the different decay modes under study, the final result is compared with pQCD calculations and with the outcome of event generators. The \Lcplus/\Dzero baryon-to-meson ratio is discussed in~\secref{sec:lcdzero}, and the results in \pp and \pPb collisions are compared with previous measurements in different collision systems and at different centre-of-mass energies, and compared with expectations from Monte Carlo \pp event generators.
Finally, in~\secref{sec:RpPb} the nuclear mo\-dification factor \RpPb is computed and compared with the results for D mesons and the predictions from models including cold-nuclear-matter and hot-medium effects.

\subsection{Prompt \Lcplus~production cross section}
\label{sec:xsecResult}

\begin{figure}[t!]
\centering
\includegraphics[width=1.0\textwidth]{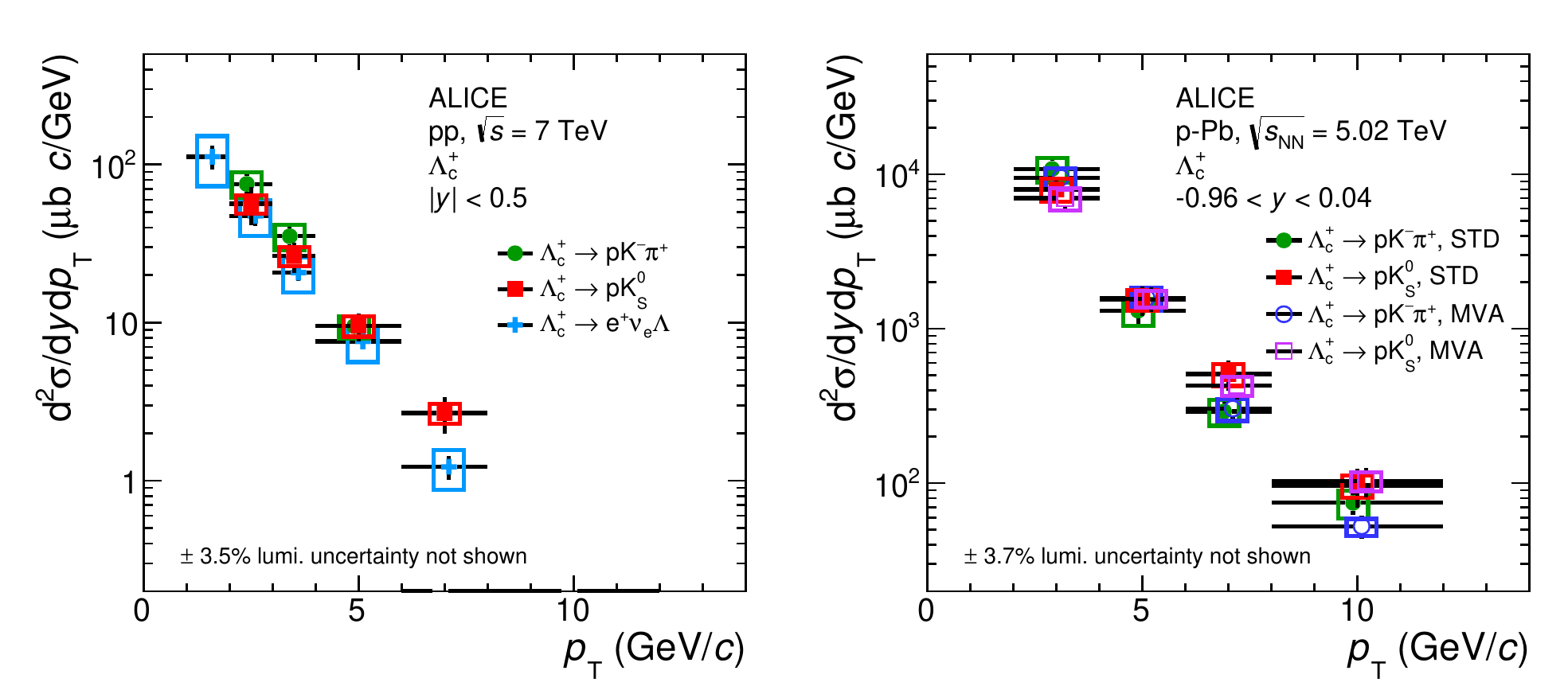}	
\caption{Prompt \Lcplus~baryon $\pt$-differential production cross section in \pp collisions at \sqrts $ = 7$~\tev in the transverse momentum interval $1 < \pt < 8$~\gevc (left) and in \pPb collisions at \sqrtsNN $= 5.02$ \tev in the transverse momentum interval $2 < \pt < 12$~\gevc (right). The statistical uncertainties are shown as error bars and the systematic uncertainties are shown as boxes. The markers for different analyses are shifted with respect to
the centre of the bin to improve visibility.}
\label{fig:CrossSection}
\end{figure}

Figure \ref{fig:CrossSection} (left) shows 
the \pt-differential cross section of prompt \Lcplus\ baryons in $\lvert y\rvert < 0.5$ in \pp collisions at $\sqrts=7~\tev$ as measured in the decay channels \LctopKpi, \LctopKzS and \LctoenuLambda (averaged with the corresponding charge conjugates).
Figure \ref{fig:CrossSection} (right) shows the \pt-differential cross section of prompt \Lcplus~in $-0.96 < y < 0.04$ in \pPb collisions at \sqrtsNN = 5.02 \tev in the decay channels \LctopKpi and \LctopKzS. In this and following figures
the marker is placed at the centre of the \pT interval unless differently specified, the horizontal bar spans the width of the \pT interval, the vertical error bar is the statistical uncertainty and the box is the systematic uncertainty.

For both collision systems, the cross sections measured from the different decay channels and analysis methods are compatible within statistical and uncorrelated systematic uncertainties, which include the uncertainty on the respective branching ratios. The largest discrepancy is observed in the \pt interval $6<\pT<8$~\gevc in \pp collisions between the \LctopKzS decay and the semileptonic decay channel, which differ by 1.7$\sigma$ after adding in quadrature statistical and uncorrelated systematic uncertainties.

To obtain a more precise determination of the cross section in each collision system, these results were averaged together, taking into account the correlation between the statistical and systematic uncertainties.
In the hadronic analyses (\LctopKpi and \LctopKzS), the sources of systematic uncertainty assumed to be uncorrelated between different decay channels are those due to the raw-yield extraction, the \Lc selection, and the PID efficiency. The sources assumed to be correlated are those due to the tracking efficiency, the generated \pt shape of the \Lc in simulation, the beauty feed-down, and the luminosity. The branching ratio uncertainties were treated as partially correlated among the hadronic decay modes, as indicated in~\cite{Olive:2016xmv}.

For the semileptonic analysis there are sources of systematic uncertainties 
that are correlated with other sources in the hadronic decay channel. In these cases, the systematic uncertainties were assumed to be fully correlated.
The uncertainties due to the reconstruction of the electron and the $\Lambda$ and the acceptance of the 
e$\Lambda$ pair are assumed to be correlated with the tracking efficiency contribution in the hadronic decay modes. The uncertainties due to the generated \pt shape of the \Lc in simulation are assumed to be correlated, as well as the contribution from the \Lb feed-down.
Other sources, including the uncertainty coming from the cuts on the e$\Lambda$ pairs, from the wrong-sign subtraction, the $\Xi_{\rm c}^{0,+}$ feed-down, the unfolding, the selections on the $\Lambda$ decay topology,  the electron identification, and the branching ratio are assumed to be fully uncorrelated between the results from the semileptonic and hadronic decay modes. 

To average the different decay channels in \pp collisions, where all measurements are statistically uncorrelated, the cross section from each decay channel was given a weight
corresponding to the inverse of the quadratic sum of the relative statistical and uncorrelated systematic uncertainties, also taking into account the partial correlation in the branching ratios, following the approach in~\cite{Avery:1996ss}.

In the case of the analyses in \pPb collisions, the cross sections in the two hadronic decay channels were measured with two different approaches, 
namely the standard cut method and the MVA method.
A high degree of correlation exists between the analyses within the same decay channel, so the statistical 
uncertainty between analyses within the same decay channel was treated as fully correlated.
The systematic uncertainty due to the yield extraction was assumed to be
uncorrelated among different analyses, while all other sources of systematic uncertainty were treated as correlated.
The statistically-correlated analyses are averaged using the relative 
uncorrelated systematic uncertainties as weights.

\begin{figure}[t!]
	\centering
	\includegraphics[width=0.45\textwidth]{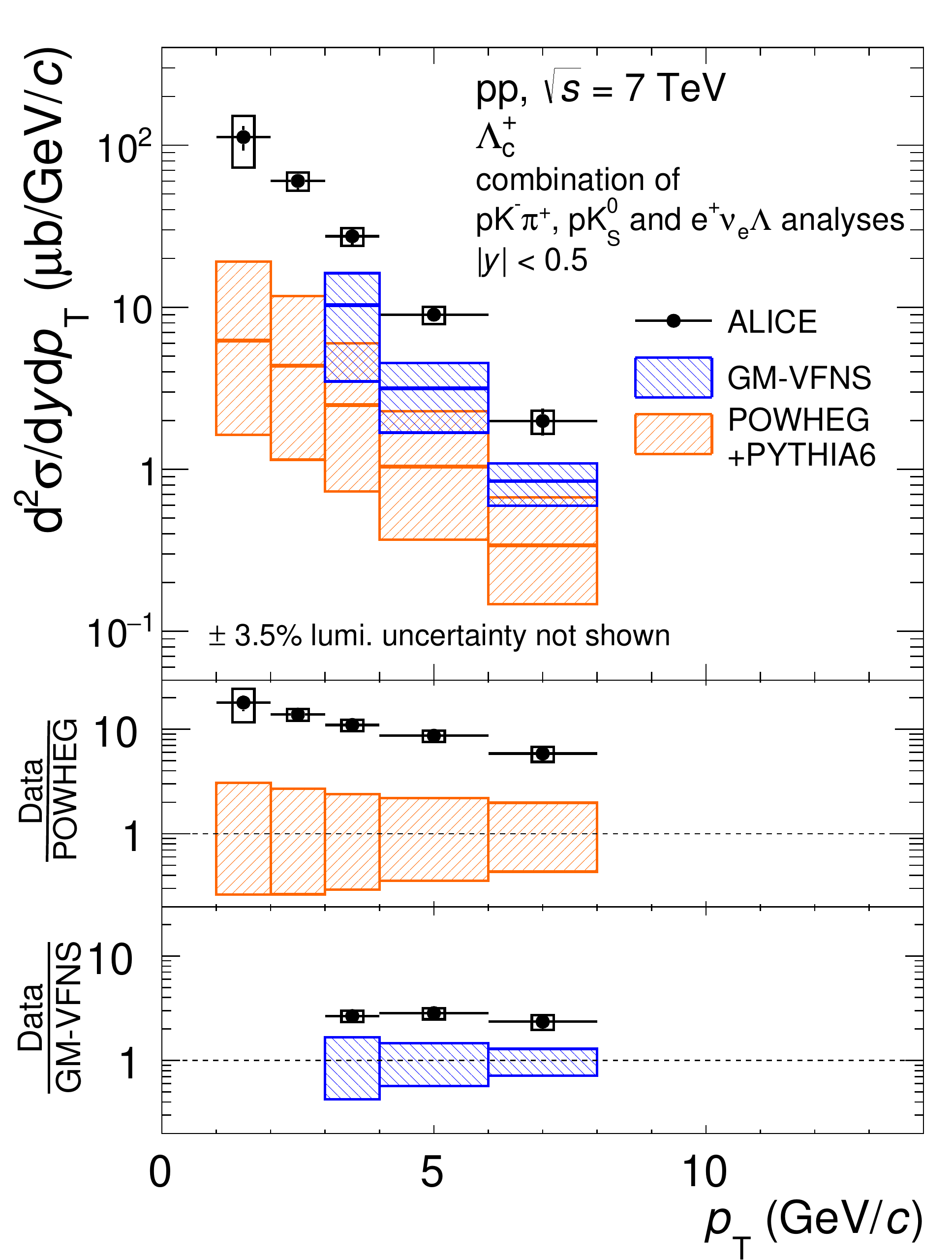} 	
	\hspace{1cm}
        \includegraphics[width=0.45\textwidth]{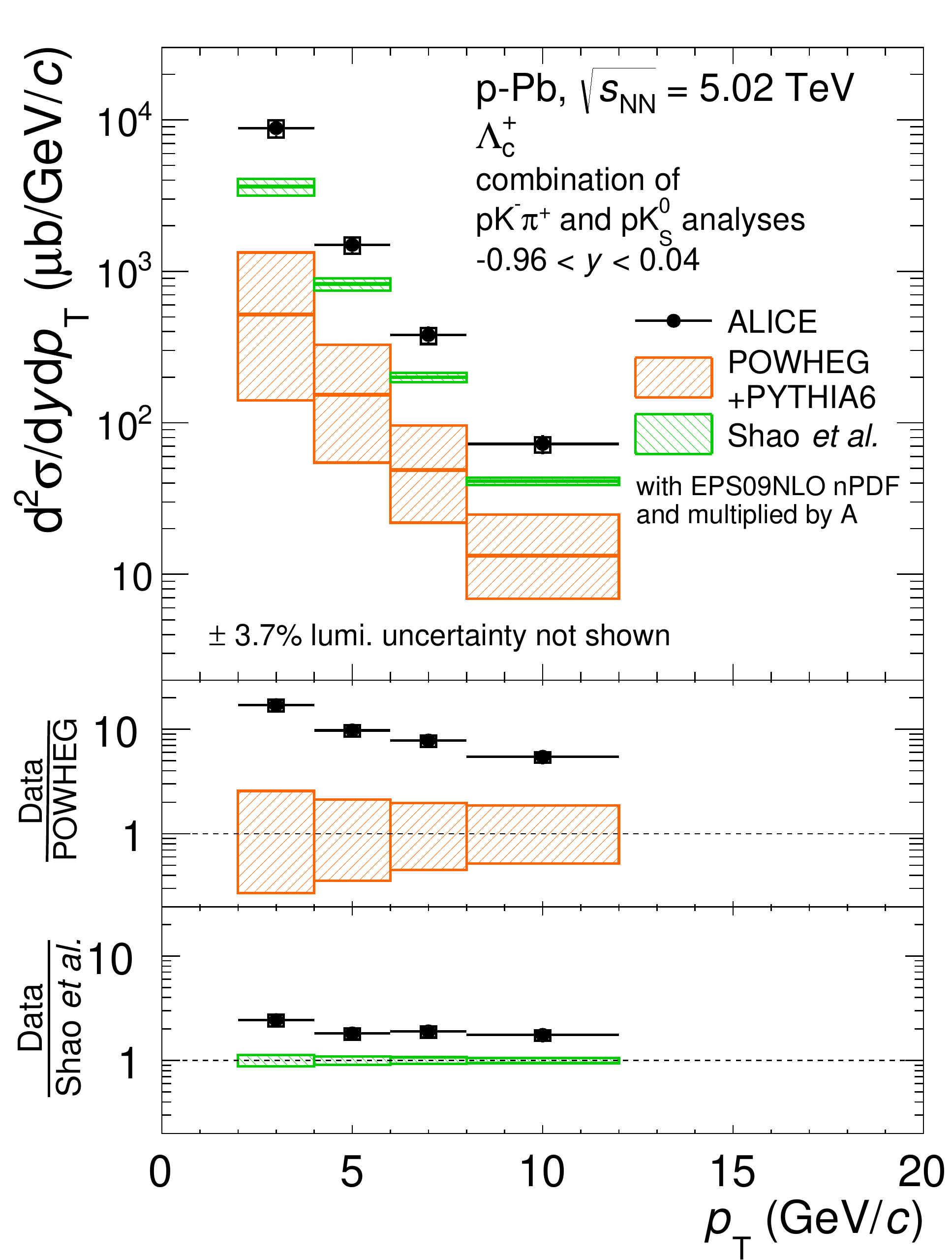} 	
	\caption{Prompt \Lcplus~baryon $\pt$-differential cross section (average among different decay modes and analyses) in \pp collisions at \sqrts $ = 7$ \tev in the transverse momentum interval $1 < \pt < 8$~\gevc (left) and in \pPb collisions at \sqrtsNN $= 5.02$ \tev in the transverse momentum interval $2 < \pt < 12$~\gevc (right). The statistical uncertainties are shown as error bars and the systematic uncertainties are shown as boxes. See text for details of the procedure to average the different decay channel measurements reported in Fig.~\ref{fig:CrossSection}. Comparisons with GM-VFNS calculations~\cite{Kniehl:2005mk,Kniehl:2012ti}, {\sc powheg} event generator~\cite{Frixione:2007nw} and with Lansberg and Shao predictions~\cite{Lansberg:2016deg} for \pPb (see text for details) are also shown.}
	\label{fig:AverageCrossSection}
\end{figure}

Figure~\ref{fig:AverageCrossSection} shows the results of the \pt-differential production cross section of prompt \Lcplus~baryons in \pp and in \pPb collisions obtained with the averaging procedure described above.
In Fig.~\ref{fig:AverageCrossSection} (left) our measurement in pp collisions is compared with GM-VFNS perturbative QCD calculations~\cite{Kniehl:2005mk,Kniehl:2012ti} and with the results of the {\sc powheg} event 
generator~\cite{Frixione:2007nw}. GM-VFNS has predictions for the \Lc baryon for \pT$>3$ \gevc and the calculations were performed using CTEQ 6.6~\cite{Nadolsky:2008zw} parameterisations of the PDFs,
assuming the charm-quark mass $m_{\rm c}$ = 1.5~\gevcc, and with the fragmentation function and fractions tuned on \ee collision data, which results in a fragmentation fraction value $f({\rm c} \rightarrow \Lc)$ = 0.061~\cite{Kniehl:2006mw}. For the {\sc powheg} calculation starting at $\pT$ = 1~\gevc, the {\sc powheg-box} package~\cite{Alioli:2010xd} was used for the NLO calculations and interfaced with {\sc pythia \small6.4.25} for the parton shower simulation and hadronisation. The {\sc powheg} calculations were performed using CT10nlo~\cite{Lai:2010vv} parameterisations of the PDF and $m_{\rm c}$ = 1.5 \gevcc.
The uncertainties shown are the envelope of the predictions obtained varying the factorisation and renormalisation scales as proposed in~\cite{Cacciari:2012ny}. 
The GM-VFNS predictions underestimate the measured cross section, which is on average higher by a factor 2.5 than the central value of the perturbative QCD calculation, as it can be seen in the bottom panel of Fig. \ref{fig:AverageCrossSection}.
Moreover, {\sc powheg} underpredicts the measured cross section by a factor of 18 (4) at low (high) \pt. 
However both GM-VFNS and {\sc powheg} describe the measured D-meson cross sections at central rapidities~\cite{Klasen:2014dba,Acharya:2017jgo} and GM-VFNS describes the \Lc cross section at forward rapidities~\cite{Aaij:2013mga}. 
It is noted that the fragmentation functions used in these calculations were derived from \ee collision data, and thus the underestimation of the data by GM-VFNS and {\sc powheg} might hint at a violation of the universality of the fragmentation functions. This possibility is for example discussed in \cite{Fischer:2016zzs} considering data in the light flavour sector. 

In Fig.~\ref{fig:AverageCrossSection} (right) the \Lc cross section in \pPb collisions is compared with the cross section obtained 
with a calculation based on {\sc powheg} using CT10nlo PDF with nuclear modification from EPS09NLO, scaled by the mass number of lead ($A$ = 208). This calculation for \pPb collisions underpredicts the measured values by a similar amount as observed in \pp collisions. The \Lc cross section is also compared with a calculation~\cite{Lansberg:2016deg}, based on a data-driven modelling of the scattering at the partonic level, specifically designed to evaluate the impact of the nuclear modification of the gluon density on heavy-flavor hadrons. The tool is based on the HELAC-Onia package~\cite{Shao:2015vga,Shao:2012iz}, originally developed for heavy-quarkonium studies, recently extended to heavy-flavor mesons and baryons. Differently from other calculations shown in Fig.~\ref{fig:AverageCrossSection}, this is therefore a prediction for \pPb collisions based on \pp data. Specifically the authors constrained their parameterisation of the cross section to the LHCb measurements of \Lc production in \pp collisions at \sqrts = 7~TeV and $2<y<4.5$~\cite{Aaij:2013mga} and they folded it with the nuclear modification of the PDFs from EPS09NLO. This model underpredicts our measurement by a factor two.

\subsection{\Lcplus/\Dzero baryon-to-meson ratio}
\label{sec:lcdzero}

The \Lcplus/\Dzero production ratio is sensitive to hadronisation mechanisms in the charm sector. For the \Dzero cross section we use the ALICE measurements~\cite{Acharya:2017jgo,Adam:2016ich}.
The \Lcplus/\Dzero ratio is computed by integrating the \pT-differential cross sections of \Lc and \Dzero (both obtained as an average of particles and anti-particles) over their common \pT interval, namely $1<\pT<8$ \gevc for \pp collisions and $2<\pT<12$ \gevc for \pPb collisions.
In the integration, the systematic uncertainty due to the raw-yield extraction in the hadronic decay analyses \LctopKpi and \LctopKzS were assumed to be fully uncorrelated between \pt intervals, and the rest of the uncertainty sources were assumed to be fully correlated between \pt intervals. 
In the \Lcplus/\Dzero ratio, the uncertainties due to the tracking efficiency, the beauty feed-down, and the luminosity were assumed to be fully correlated between the \Lcplus~and \Dzero cross sections, and all other sources were assumed to be fully uncorrelated.
The resulting baryon-to-meson ratio \Lcplus/\Dzero measured in \pp collisions at \sqrts=7 TeV, $|y|<0.5$, and $1<\pt<8$ \gevc is

\begin{equation}
\left(\frac{\Lcplus}{\Dzero}\right)_{\pp} = 0.543~ \pm ~0.061 {\rm ~(stat)~} \pm 0.160 {\rm ~(syst)}. 
\end{equation}

\begin{figure}[t!]
	\centering
	\includegraphics[width=0.47\textwidth]{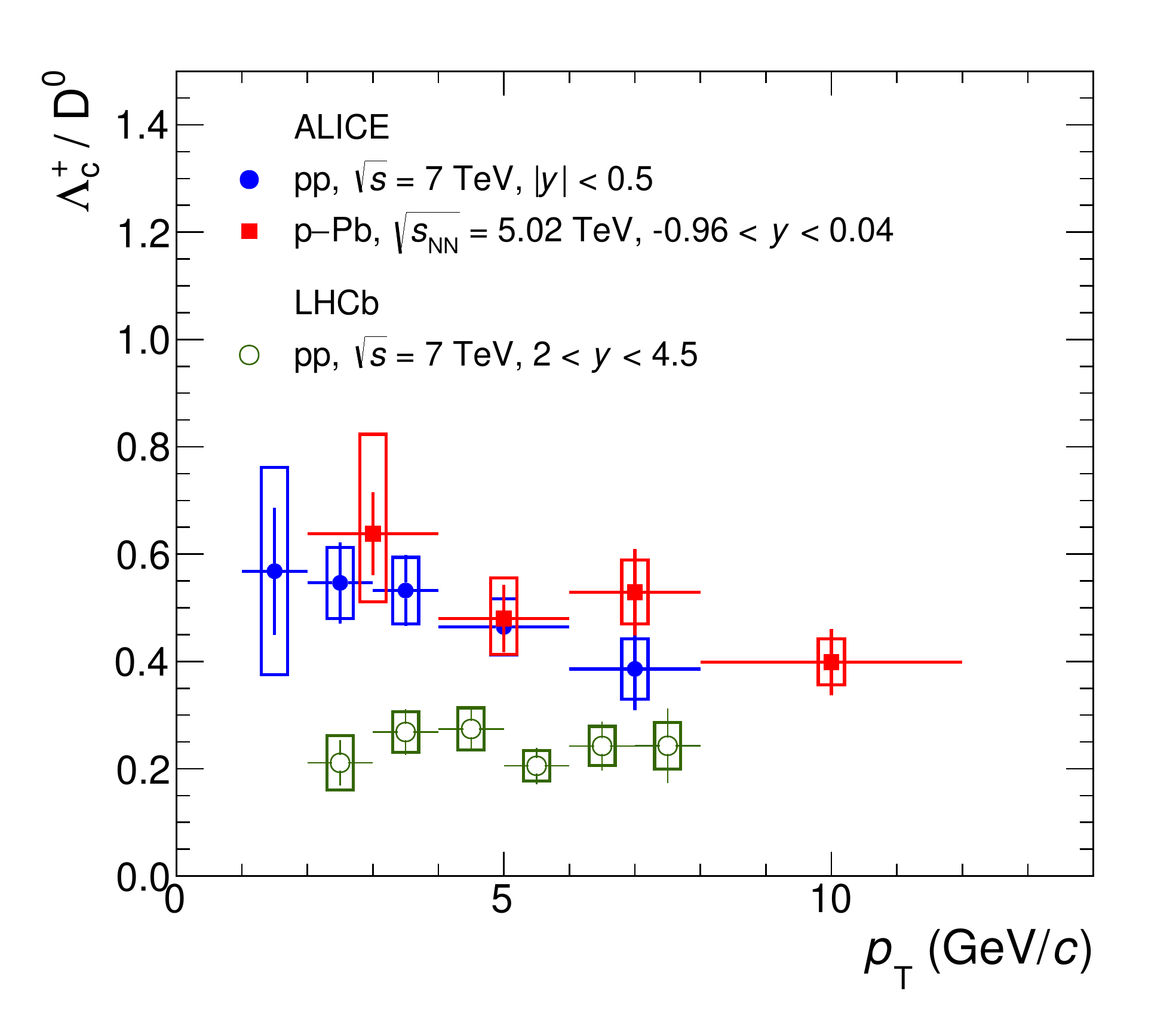}
	\includegraphics[width=0.47\textwidth]{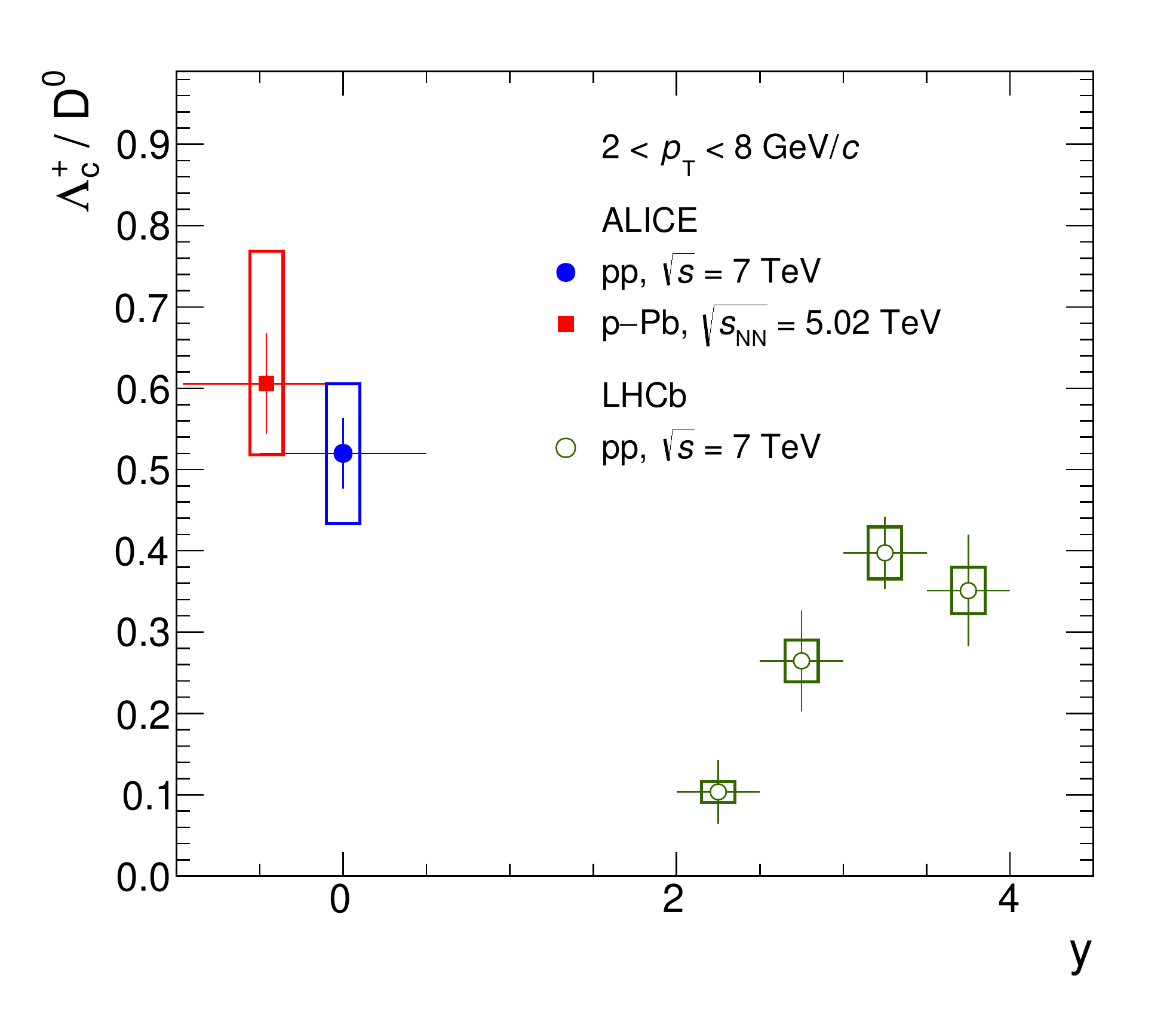}
	\caption{The \Lcplus/\DZero ratio measured in \pp and \pPb collisions by ALICE, compared with the LHCb measurement~\cite{Aaij:2013mga,LHCbNote} as a function of \pt (left) and as a function of $y$ for $2 < \pt < 8$ \gevc (right).}
	\label{fig:LHCbComparison}
\end{figure}

In \pPb collisions at \sqrtsNN=5.02 TeV, $-0.96<y<0.04$, and $2<\pT<12$ \gevc the measured baryon-to-meson ratio is

\begin{equation}
\left(\frac{\Lcplus}{\Dzero}\right)_{\pPb} = 0.602~ \pm ~0.060 {\rm ~(stat)~} \substack{+0.159\\ -0.087} {\rm ~(syst)},
\end{equation}
and is compatible within uncertainty with that measured in \pp collisions.
A list of existing measurements of the \Lcplus/\Dzero ratio in different collision systems and kinematic ranges is reported in Tab.~\ref{tab:lcd0ratio}.
In Fig.~\ref{fig:LHCbComparison}, the measured \Lcplus/\DZero ratio in \pp and \pPb collisions is presented as a function of \pt (left panel) and rapidity (right panel) and compared with the LHCb measurement in \pp collisions, with values derived by the LHCb Collaboration~\cite{LHCbNote} from their published result~\cite{Aaij:2013mga}.

\begin{table}[h!]
\small
\begin{center}
\def\arraystretch{1.4}\tabcolsep=6pt  
   \begin{tabular}{lrccr} 
\toprule
 & \Lcplus/\Dzero $\pm$ stat. $\pm$ syst. & System & \sqrts~(GeV)& Notes \\
\midrule
CLEO~\cite{Avery:1990bc} & $0.119 \pm 0.021 \pm 0.019$ & ee & $10.55$   & \\
ARGUS~\cite{Albrecht:1988an,Albrecht:1991ss} & $0.127 \pm 0.031$ & ee & $10.55$  & \\
LEP average~\cite{Gladilin:2014tba} & $0.113 \pm 0.013 \pm 0.006$ & ee & $91.2$  & \\
\multirow{2}{*}{ZEUS DIS~\cite{Abramowicz:2010aa}} & \multirow{2}{*}{$0.124 \pm 0.034 \substack{ +0.025\\ -0.022}$} & \multirow{2}{*}{ep} & \multirow{2}{*}{$320$}   & $1 < Q^2 < 1000$ $\gev ^2$, \\
& & & & $0 < \pt < 10$ \gevc, $0.02 < y < 0.7$\\[0.2cm]	
ZEUS $\gamma$p,  & \multirow{2}{*}{$0.220 \pm 0.035 \substack{+0.027\\ -0.037} $} & \multirow{2}{*}{ep} & \multirow{2}{*}{$320$} & $130 < W < 300$ \gev, $Q^2 < 1$ $\gev^2$, \\
HERA I~\cite{Chekanov:2005mm} & & & & $\pt > 3.8$ \gevc, $|\eta| < 1.6$\\[0.2cm]
ZEUS $\gamma$p,  & \multirow{2}{*}{$0.107 \pm 0.018 \substack{+0.009\\ -0.014} $} & \multirow{2}{*}{ep} & \multirow{2}{*}{$320$} & $130 < W < 300$ \gev, $Q^2 < 1$ $\gev^2$,\\
HERA II~\cite{Abramowicz:2013eja} & & & & $\pt > 3.8$ \gevc, $|\eta| < 1.6$\\[0.2cm]
\midrule
\end{tabular}
\end{center}
\caption{Comparison of the \Lcplus/\Dzero ratio as measured in \ee and \ep collision systems and at different centre-of-mass energies. Statistical and systematic uncertainties are reported (from references~\cite{Albrecht:1988an,Albrecht:1991ss} it was not possible to separate systematics and statistical uncertainties). See text for details about how the central values and quoted uncertainties were obtained. When indicated, the rapidity range refers to the centre-of-mass frame.}
\label{tab:lcd0ratio}  
\end{table}

For the measurements in \ee and \ep collisions reported in Tab.~\ref{tab:lcd0ratio} and for the LHCb results reported in Fig.~\ref{fig:LHCbComparison} the central values were multiplied by a correction factor that takes into account the most recent values of the BR of the \LctopKpi and 
\DtoKpi decays~\cite{Olive:2016xmv}. Wherever the systematic uncertainties for the bran\-ching ratios were quoted separately, they were updated according to the most recent values. Luminosity systematic uncertainties that cancel out in the ratio were not considered.
The \Lcplus/\Dzero ratio was obtained, when available, from the ratio of the measured fragmentation fractions $f({\rm c}\rightarrow\Lc)$ and $f({\rm c}\rightarrow\Dzero)$.

As shown in the table, a comparison is not straightforward given the different energy scales, the different collision systems and the fact that the extrapolation in the phase space down to \pt = 0 was done for only a fraction of all measurements.
The ratio \Lcplus/\Dzero can depend on the \pt interval in which it is evaluated because of the possible differences in the fragmentation functions of charm quarks into baryons and mesons, which would result in different momentum distributions of \Lc baryons as compared to \Dzero mesons.
The results reported in this paper for the \Lcplus/\Dzero ratio are higher than previous measurements carried out in \ee and \ep collisions, and at lower centre-of-mass energies, where proposed mechanisms expected to enhance baryon production should play a negligible role as discussed in Sec.~\ref{sec: Introduction}. In the beauty sector a difference
in the fragmentation fraction $f({\rm b}\rightarrow\Lb)$ has been reported, with larger values observed in \ppbar and \pp collisions, respectively at Tevatron~\cite{Aaltonen:2008zd} and at the LHC~\cite{Aaij:2015fea}, with respect to \ee collisions at LEP~\cite{Gladilin:2014tba}.

As shown in Fig.~\ref{fig:LHCbComparison} the ratios measured by ALICE in \pp and \pPb collisions at mid-rapidity are compatible, both as a function of \pt and \pt-integrated, within uncertainties. The LHCb result in rapidity intervals suggests a decreasing trend towards mid-rapidity (influencing in turn the rapidity-averaged values reported in Fig.~\ref{fig:LHCbComparison}~(left)) that is not consistent with the ALICE result despite the large uncertainties. Such a trend is not reported by LHCb in their recent preliminary result in \pPb collisions~\cite{LHCb:2017rvh}.
Although the ALICE result seems to decrease with increasing transverse momentum, a firm conclusion cannot be drawn as to whether the observed difference between the \Lcplus/\Dzero ratios at forward and mid-rapidity is significantly \pt-dependent.

\begin{figure}[ht!]
	\centering
\includegraphics[width=0.47\textwidth]{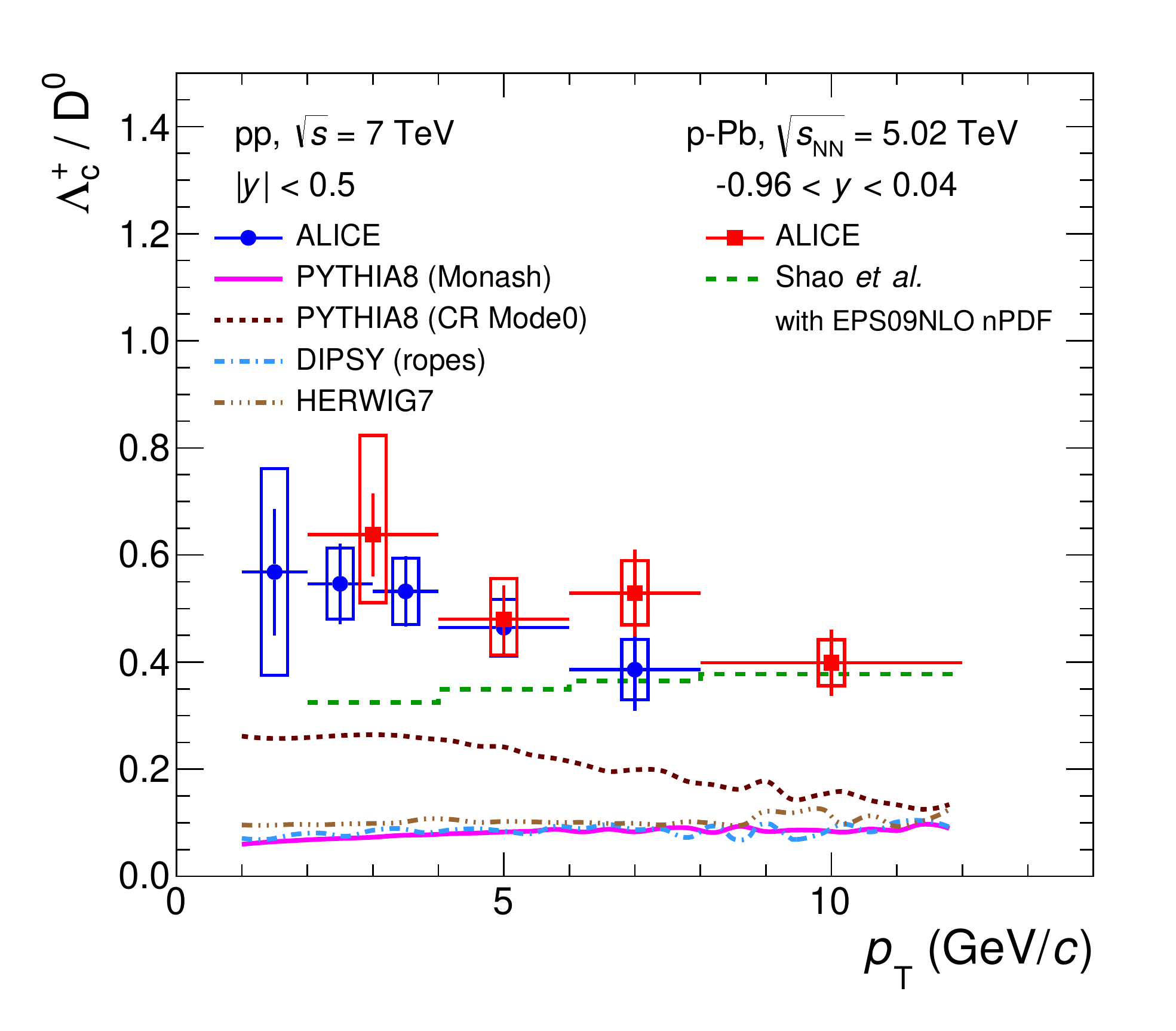}	
      \includegraphics[width=0.47\textwidth]{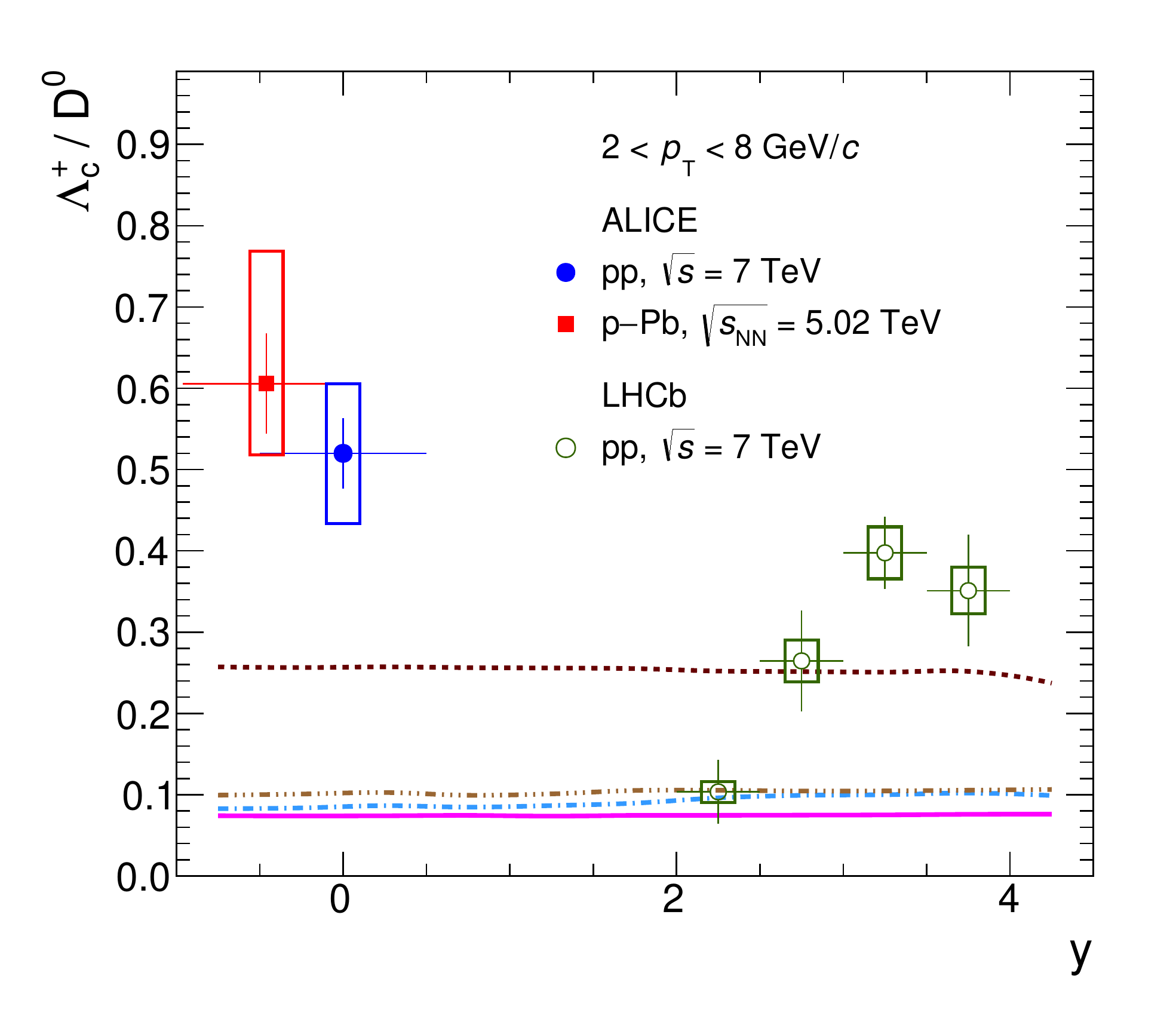}
	\caption{The \Lcplus/\DZero ratio measured in \pp and \pPb collisions by ALICE as a function of \pt (left) and as a function of $y$ for $2 < \pt < 8$ \gevc (right). The measurements from \pp collisions are compared with different event generators (quoted tunes for {\sc pythia} and {\sc dipsy} taken respectively from~\cite{Christiansen:2015yqa} and~\cite{Bierlich:2015rha}). The \pPb measurement as a function of \pT is also compared with calculations from Lansberg and Shao~\cite{Lansberg:2016deg}. The predictions from event generators as a function of $y$ are also compared with the LHCb measurement~\cite{Aaij:2013mga,LHCbNote}.}
	\label{fig:lcd0ratio}
\end{figure}

Figure ~\ref{fig:lcd0ratio} compares the \Lcplus/\Dzero ratio as a function of \pT (left panel) and rapidity (right panel) in \pp and \pPb collisions with predictions obtained from Monte Carlo \pp event generators, 
namely {\sc pythia\small8} with Monash tune and with another tune~\cite{Christiansen:2015yqa} that includes a model of string formation beyond the leading-colour approximation, {\sc dipsy} with rope parameters taken from~\cite{Bierlich:2015rha}, and {\sc Herwig7} 
which uses a cluster hadronisation mechanism.
As for the cross section calculations described in Sec.~\ref{sec:xsecResult}, fragmentation parameters for these predictions are derived from e$^+$e$^-$ collision data.
The enhanced colour reconnection mechanisms enabled in {\sc pythia\small8} increase the baryon-to-meson
ratio in the charm sector, bringing the prediction closer to the data at mid-rapidity.
The {\sc dipsy} generator with a rope configuration, which is expected to increase the baryon-to-meson ratio, instead predicts values similar
to those from {\sc pythia\small8} with Monash tune, which are lower than the values in \ee and \ep collisions as reported in Tab.~\ref{tab:lcd0ratio}. Similar predictions were obtained with {\sc Herwig7}.
The \pPb measurement is compared then in Fig.~\ref{fig:lcd0ratio}~(left) with the calculations from Lansberg and Shao~\cite{Lansberg:2016deg} for \pPb, with \Lcplus~ and \Dzero cross section obtained through a parameterisation of \pp data and using EPS09NLO nuclear modification factors. Among the different predictions this calculation is the closest to data. 
Finally, all models predict a flat rapidity dependence which does not describe the trend observed at forward rapidity. The preliminary result from LHCb in \pPb collisions~\cite{LHCb:2017rvh} also shows a flat rapidity
dependence in the $1.5<y<4$ interval.

\subsection{\Lc-baryon nuclear modification factor in \pPb collisions at \sqrtsNN = 5.02 TeV}
\label{sec:RpPb}

The nuclear modification factor $\RpPb$ of \Lc baryons was calculated from the results presented in~\secref{sec:xsecResult}
by dividing the \pt-differential prompt production cross section in \pPb collisions at \sqrtsNN= 5.02~\tev by that in \pp collisions corrected for the different centre-of-mass energy and rapidity coverage of the \pp and \pPb measurements and multiplied by the mass number $A$ = 208.

In particular, the cross section in \pp collisions measured at \sqrts = 7~\tev and $|y|<0.5$
was scaled in each \pT interval to \sqrts = 5.02~\tev and ${-0.96<y<0.04}$
using a factor $f^{\sqrts,y}_{\rm FONLL}$ calculated with
FONLL perturbative QCD calculations~\cite{Cacciari:2012ny}, following a similar procedure to the D-meson \RpPb measurement~\cite{Adam:2016ich}:

\begin{equation}	
  R_{\mathrm{pPb}} = \frac{1}{A} \frac{\mathrm{d}\sigma^{\rm 5 TeV}_{\mathrm{pPb}}/\mathrm{d}\pt}
                             {f^{\sqrts,y}_{\rm FONLL}(\pt) \cdot \mathrm{d}\sigma^{\rm 7 TeV}_{\mathrm{pp}}/\mathrm{d}\pt} \qquad  \left( f^{\sqrt{s},y}_{\rm FONLL}(p_{\rm T}) = \frac{\mathrm{d}\sigma^{\rm 5 TeV}_{\rm FONLL}/\mathrm{d}\pt}{\mathrm{d}\sigma^{\rm 7 TeV}_{\rm FONLL}/\mathrm{d}\pt}\right),
\label{eq:RpPbScaled}
\end{equation}

with FONLL cross sections calculated at 7~\tev in $|y|<0.5$, and at 5.02~\tev in ${-0.96<y<0.04}$. The uncertainties on the scaling factor are calculated by consistently varying the charm-quark mass, the PDF, and the factorisation and renormalisation scales in the calculations at the two energies.
  
The fragmentation function of charm quarks into \Lc baryons is not well known. However, it has been ve\-rified that changing the fragmentation function does not change the scaling factor significantly: the $f^{\sqrts,y}_{\rm FONLL}$ values obtained from FONLL calculations for \Dzero, \Dplus and \Dstar vary by less than $1\%$. For this reason, the \Dplus production cross section ratio from FONLL was chosen for the central values of $f^{\sqrts,y}_{\rm FONLL}(\pt)$, and the uncertainty was estimated by varying the fragmentation function.
The bare c-quark cross section from FONLL defines the upper uncertainty of the scaling factor, as the ``hardest'' fragmentation case, where it is assumed that all the momentum of the c quark is carried by the \Lc. The c-quark cross section from FONLL, convolved with a fragmentation function modelled using the Peterson parameterisation~\cite{Peterson:1982ak} with $\epsilon$ = 0.1, defines the lower uncertainty of the scaling factor as the ``softest'' case. For both limits, the associated uncertainties from FONLL were included.
These two scenarios were chosen to encompass the values reported by the PDG review for charm- and beauty-quark fragmentation for different models of hard radiation~\cite{Olive:2016xmv}. It has also been verified that the \Lcplus / \Dzero ratio obtained using these fragmentation scenarios for the \Lcplus~ and the \Dzero cross section from FONLL is compatible with the measured \Lcplus/\Dzero ratio.
The uncertainty on the scaling factor varies from $^{+13}_{-5}\%$ in the $\pt$ interval 2-4 \gevc to $^{+6}_{-4}\%$ in the $\pt$ interval 6-8 \gevc.

\begin{figure}[t!]
	\centering
	\includegraphics[width=0.85\textwidth]{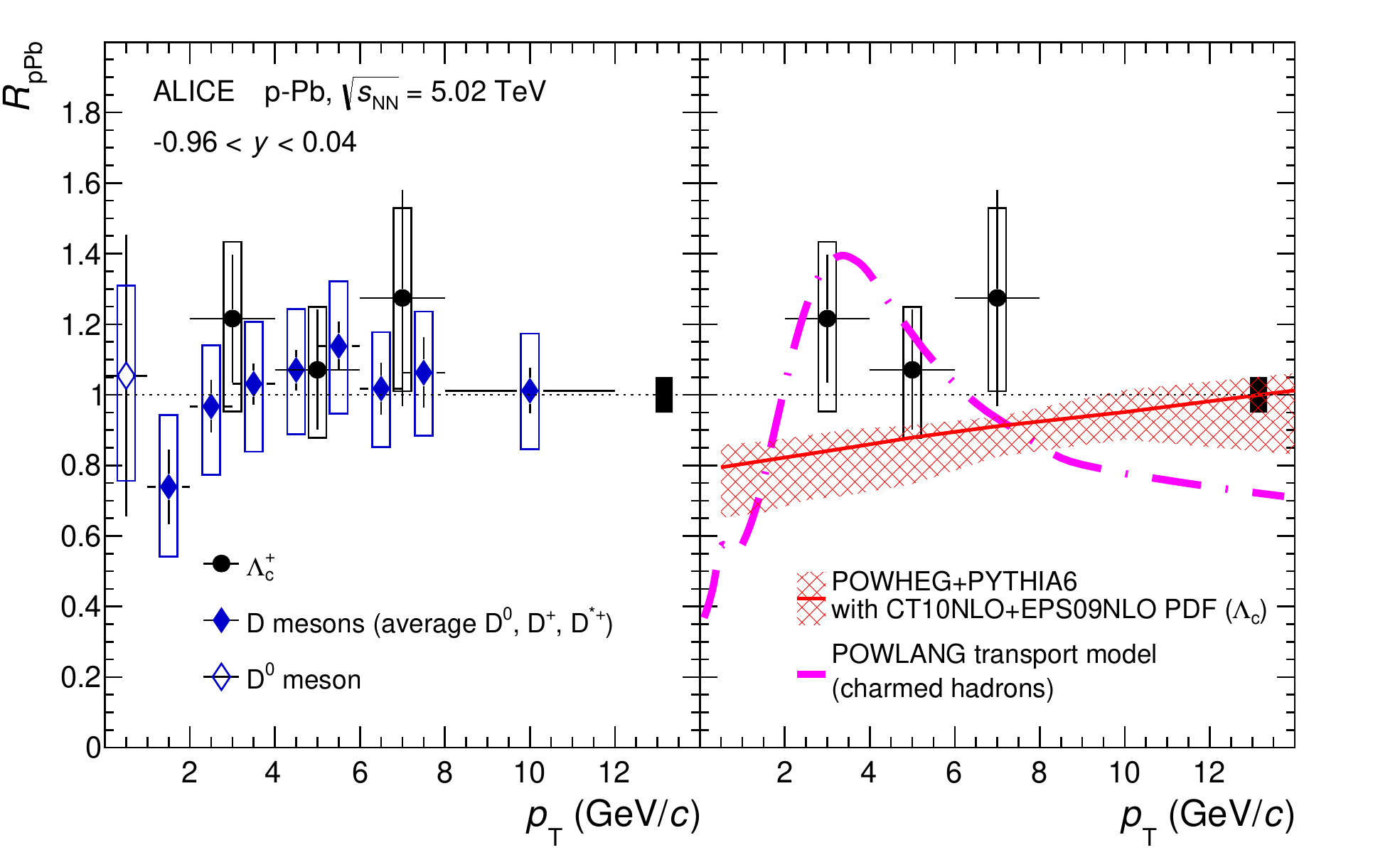}
	\caption{The nuclear modification factor \RpPb of prompt \Lcplus~baryons  in \pPb collisions at \sqrtsNN = 5.02 TeV as a function of \pT compared to that of D mesons (average of \Dzero, \Dplus and \Dstar in the range $1 < \pt < 12$~\gevc, and \Dzero in the range $0 < \pt < 1$~\gevc)~\cite{Adam:2016ich} (left panel) and to model calculations (right panel). The predictions for the comparison are the \Lcplus~\RpPb~from the {\sc powheg} event generator~\cite{Frixione:2007nw} with EPS09NLO parameterisation of the nuclear modification of the PDFs~\cite{Eskola:2009uj} and the charm-hadron \RpPb from the POWLANG transport model~\cite{Beraudo:2015wsd} assuming a QGP is formed in \pPb collisions.}
	\label{fig:RpPb}
\end{figure}

For the propagation of the uncertainties in the \RpPb computation, the beauty feed-down uncertainties are considered fully correlated between the \pp and \pPb cross sections and the branching ratio uncertainties are considered
partially correlated due to the different decay modes considered in the two collision sy\-stems, while
all the other systematic uncertainties are treated as uncorrelated. 
The uncertainty due to the \sqrts and rapidity scaling of the \pp reference
was added in quadrature to the aforementioned sources. The luminosity uncertainties were treated as fully uncorrelated.

Figure~\ref{fig:RpPb} (left) shows the \Lc-baryon nuclear modification factor \RpPb in the range $2 < \pt < 8$ \gevc.
The result is compatible with unity within the large statistical and systematic uncertainties, and is consistent with the D-meson  
\RpPb~\cite{Adam:2016ich}, which is shown in the same figure. Predictions for the \RpPb for \Lc baryons from the {\sc powheg} event generator with {\sc pythia} 
parton shower~\cite{Frixione:2007nw} and EPS09NLO parameterisation of nuclear modification of the PDFs~\cite{Eskola:2009uj} are presented in the 
right panel of Fig.~\ref{fig:RpPb}.
In the same panel, the calculations for the charmed-hadron nuclear modification factor from the POWLANG model~\cite{Beraudo:2015wsd}, which assumes 
that also in \pPb collisions at LHC energies a hot deconfined medium is formed, 
is superimposed.
The POWLANG model utilises the Langevin approach to compute the transport of heavy quarks through an expanding QGP described by relativistic viscous hydrodynamics, but it does not include any specific mechanism to modify hadronisation, such as coalescence, that could lead to a baryon enhancement. 
This transport model predicts a deviation of \RpPb from unity which is about 20-40\% at low and intermediate momentum (\pT$<5$~\gevc).
The precision achieved with the current measurement does not allow us to distinguish between calculations with and without hot medium effects.

\section{Conclusions}
\label{sec: Conclusions}
We measured the \Lc baryon production in \pp and \pPb collisions with ALICE at the LHC using different decay channels and different analysis methods. In \pp collisions, we reported the production cross section measurement at mid-rapidity ($|y| < 0.5$) and \sqrts = 7 \tev for this baryon, while in \pPb collisions the \Lc production cross section was measured at the centre-of-mass energy per nucleon-nucleon pair ${\sqrt{s_{\textsc{NN}}}\xspace=\xspace{\rm 5.02~TeV}}$ in the centre-of-mass rapidity interval $-0.96 < y < 0.04$.
The results were reported for \pp collisions in the transverse-momentum interval $1<\pT<8$ \gevc and for \pPb collisions in $2<\pT<12$ \gevc.

The measurement of the \Lc baryon, due to its short lifetime, is challenging: the \pT-differential production cross sections were therefore obtained averaging the results from different decay channels (purely hadronic and semileptonic) and with different analysis approaches, using standard cuts, Multivariate Analysis techniques and a dedicated  procedure to subtract background pairs for the semileptonic channel. Different  PID-discriminating variables were also used. The results of all the analyses were found to be mutually consistent within uncertainties.

In the \pT interval where calculations from the GM-VFNS perturbative QCD framework are available ($3 < \pT < 8$ \gevc), the predictions underestimate the measured cross section on average by a factor of 2.5. The comparison is, however, affected by large uncertainties, in particular from the theoretical estimates. Calculations based on {\sc powheg} (available for $\pT > 1$ \gevc) with hadronisation from the {\sc pythia}  parton shower, underpredict the measured values by a factor of 18 (4) at low (high) \pt. A similar pattern is observed  comparing cross section predictions obtained with {\sc powheg} with measured values in \pPb collisions. Calculations for this collision system based on a parameterisation of existing \pp measurements for \Lc are closer to the data, even if they are still underpredicting measured values by a factor of 2.

The baryon-to-meson ratios \Lcplus/\Dzero measured in \pp and \pPb collisions are compatible within their statistical and systematic uncertainties. Our result in \pp collisions (\Lcplus/\Dzero = $0.543 \pm 0.061 \pm 0.160 $ for $1<\pT<8$ \gevc at mid-rapidity) is larger than previous measurements obtained at lower centre-of-mass energies and in different collision systems, and also higher than the results reported by LHCb at $2<y<4.5$ for $2<\pT<8$ \gevc in \pp collisions at the same centre-of-mass energy.

We also compared the measured \Lcplus/\Dzero ratio to \pp event generators that implement different hadronisation schemes. All underpredict the measured values: a better qualitative agreement with our results is obtained with {\sc pythia} tunes that include string formation beyond the leading-colour approximation, while significantly lower values are obtained with {\sc dipsy} and {\sc Herwig7}. 

Finally, a first measurement of the nuclear mo\-dification factor \RpPb~of \Lc baryons was obtained and it was found to be compatible with unity in the transverse-momentum interval $2<\pT<8$~\gevc, as well as with the \RpPb~of D mesons. The current precision of the measurement cannot constrain existing models.

When considered in their entirety, these results provide input for theoretical models based on pQCD calculations, event generators applying different hadronisation approaches and models describing CNM and/or hot-medium effects in proton--nucleus collisions. A better precision is expected to be reached with data presently collected during the LHC {\sc run} 2, reducing in particular the statistical uncertainties, and, in the future, during the LHC {\sc run} 3 and 4 following a major upgrade of the ALICE apparatus~\cite{Abelevetal:2014cna}. This set of measurements provides an initial reference for future investigation of \Lc production in \PbPb collisions where the interaction of charm quarks with the hot medium may affect its production.

\clearpage

\clearpage

\newenvironment{acknowledgement}{\relax}{\relax}
\begin{acknowledgement}
\section*{Acknowledgements}
The ALICE Collaboration would like to thank M.~Corradi and L.~Gladilin from ZEUS Collaboration for providing guidance on how to treat the different sources of uncertainties in the computation of the \Lcplus/\Dzero ratio for the ZEUS data, P.~Spradlin and V.~Vagnoni from LHCb Collaboration for providing \Lcplus/\Dzero baryon-to-meson ratio shown in Fig.~\ref{fig:LHCbComparison} in this paper, and P.~Robbe for his help to configure the {\sc EvtGen} package.

The ALICE Collaboration would like to thank all its engineers and technicians for their invaluable contributions to the construction of the experiment and the CERN accelerator teams for the outstanding performance of the LHC complex.
The ALICE Collaboration gratefully acknowledges the resources and support provided by all Grid centres and the Worldwide LHC Computing Grid (WLCG) collaboration.
The ALICE Collaboration acknowledges the following funding agencies for their support in building and running the ALICE detector:
A. I. Alikhanyan National Science Laboratory (Yerevan Physics Institute) Foundation (ANSL), State Committee of Science and World Federation of Scientists (WFS), Armenia;
Austrian Academy of Sciences and Nationalstiftung f\"{u}r Forschung, Technologie und Entwicklung, Austria;
Ministry of Communications and High Technologies, National Nuclear Research Center, Azerbaijan;
Conselho Nacional de Desenvolvimento Cient\'{\i}fico e Tecnol\'{o}gico (CNPq), Universidade Federal do Rio Grande do Sul (UFRGS), Financiadora de Estudos e Projetos (Finep) and Funda\c{c}\~{a}o de Amparo \`{a} Pesquisa do Estado de S\~{a}o Paulo (FAPESP), Brazil;
Ministry of Science \& Technology of China (MSTC), National Natural Science Foundation of China (NSFC) and Ministry of Education of China (MOEC) , China;
Ministry of Science, Education and Sport and Croatian Science Foundation, Croatia;
Ministry of Education, Youth and Sports of the Czech Republic, Czech Republic;
The Danish Council for Independent Research | Natural Sciences, the Carlsberg Foundation and Danish National Research Foundation (DNRF), Denmark;
Helsinki Institute of Physics (HIP), Finland;
Commissariat \`{a} l'Energie Atomique (CEA) and Institut National de Physique Nucl\'{e}aire et de Physique des Particules (IN2P3) and Centre National de la Recherche Scientifique (CNRS), France;
Bundesministerium f\"{u}r Bildung, Wissenschaft, Forschung und Technologie (BMBF) and GSI Helmholtzzentrum f\"{u}r Schwerionenforschung GmbH, Germany;
General Secretariat for Research and Technology, Ministry of Education, Research and Religions, Greece;
National Research, Development and Innovation Office, Hungary;
Department of Atomic Energy Government of India (DAE), Department of Science and Technology, Government of India (DST), University Grants Commission, Government of India (UGC) and Council of Scientific and Industrial Research (CSIR), India;
Indonesian Institute of Science, Indonesia;
Centro Fermi - Museo Storico della Fisica e Centro Studi e Ricerche Enrico Fermi and Istituto Nazionale di Fisica Nucleare (INFN), Italy;
Institute for Innovative Science and Technology , Nagasaki Institute of Applied Science (IIST), Japan Society for the Promotion of Science (JSPS) KAKENHI and Japanese Ministry of Education, Culture, Sports, Science and Technology (MEXT), Japan;
Consejo Nacional de Ciencia (CONACYT) y Tecnolog\'{i}a, through Fondo de Cooperaci\'{o}n Internacional en Ciencia y Tecnolog\'{i}a (FONCICYT) and Direcci\'{o}n General de Asuntos del Personal Academico (DGAPA), Mexico;
Nederlandse Organisatie voor Wetenschappelijk Onderzoek (NWO), Netherlands;
The Research Council of Norway, Norway;
Commission on Science and Technology for Sustainable Development in the South (COMSATS), Pakistan;
Pontificia Universidad Cat\'{o}lica del Per\'{u}, Peru;
Ministry of Science and Higher Education and National Science Centre, Poland;
Korea Institute of Science and Technology Information and National Research Foundation of Korea (NRF), Republic of Korea;
Ministry of Education and Scientific Research, Institute of Atomic Physics and Romanian National Agency for Science, Technology and Innovation, Romania;
Joint Institute for Nuclear Research (JINR), Ministry of Education and Science of the Russian Federation and National Research Centre Kurchatov Institute, Russia;
Ministry of Education, Science, Research and Sport of the Slovak Republic, Slovakia;
National Research Foundation of South Africa, South Africa;
Centro de Aplicaciones Tecnol\'{o}gicas y Desarrollo Nuclear (CEADEN), Cubaenerg\'{\i}a, Cuba and Centro de Investigaciones Energ\'{e}ticas, Medioambientales y Tecnol\'{o}gicas (CIEMAT), Spain;
Swedish Research Council (VR) and Knut \& Alice Wallenberg Foundation (KAW), Sweden;
European Organization for Nuclear Research, Switzerland;
National Science and Technology Development Agency (NSDTA), Suranaree University of Technology (SUT) and Office of the Higher Education Commission under NRU project of Thailand, Thailand;
Turkish Atomic Energy Agency (TAEK), Turkey;
National Academy of  Sciences of Ukraine, Ukraine;
Science and Technology Facilities Council (STFC), United Kingdom;
National Science Foundation of the United States of America (NSF) and United States Department of Energy, Office of Nuclear Physics (DOE NP), United States of America.
\end{acknowledgement}



\small 

\bibliographystyle{utphys}
\bibliography{include/bibfile}

\cleardoublepage
\normalsize 
\appendix
\section{The ALICE Collaboration}
\label{app:collab}

\begingroup
\small
\begin{flushleft}
S.~Acharya\Irefn{org138}\And 
F.T.-.~Acosta\Irefn{org22}\And 
D.~Adamov\'{a}\Irefn{org94}\And 
J.~Adolfsson\Irefn{org81}\And 
M.M.~Aggarwal\Irefn{org98}\And 
G.~Aglieri Rinella\Irefn{org36}\And 
M.~Agnello\Irefn{org33}\And 
N.~Agrawal\Irefn{org48}\And 
Z.~Ahammed\Irefn{org138}\And 
S.U.~Ahn\Irefn{org77}\And 
S.~Aiola\Irefn{org143}
% ...
\And
\renewcommand\labelenumi{\textsuperscript{\theenumi}~}

\section*{Affiliation notes}
\renewcommand\theenumi{\roman{enumi}}
\begin{Authlist}
\item \Adef{org*}Deceased
\item \Adef{orgI}Dipartimento DET del Politecnico di Torino, Turin, Italy
\item \Adef{orgII}M.V. Lomonosov Moscow State University, D.V. Skobeltsyn Institute of Nuclear, Physics, Moscow, Russia
\item \Adef{orgIII}Department of Applied Physics, Aligarh Muslim University, Aligarh, India
\item \Adef{orgIV}Institute of Theoretical Physics, University of Wroclaw, Poland
\end{Authlist}

\section*{Collaboration Institutes}
\renewcommand\theenumi{\arabic{enumi}~}
\begin{Authlist}
\item \Idef{org1}A.I. Alikhanyan National Science Laboratory (Yerevan Physics Institute) Foundation, Yerevan, Armenia
\item \Idef{org2}Benem\'{e}rita Universidad Aut\'{o}noma de Puebla, Puebla, Mexico
\item \Idef{org3}Bogolyubov Institute for Theoretical Physics, National Academy of Sciences of Ukraine, Kiev, Ukraine
\item \Idef{org4}Bose Institute, Department of Physics  and Centre for Astroparticle Physics and Space Science (CAPSS), Kolkata, India
\item \Idef{org5}Budker Institute for Nuclear Physics, Novosibirsk, Russia
\item \Idef{org6}California Polytechnic State University, San Luis Obispo, California, United States
\item \Idef{org7}Central China Normal University, Wuhan, China
\item \Idef{org8}Centre de Calcul de l'IN2P3, Villeurbanne, Lyon, France
\item \Idef{org9}Centro de Aplicaciones Tecnol\'{o}gicas y Desarrollo Nuclear (CEADEN), Havana, Cuba
\item \Idef{org10}Centro de Investigaci\'{o}n y de Estudios Avanzados (CINVESTAV), Mexico City and M\'{e}rida, Mexico
\item \Idef{org11}Centro Fermi - Museo Storico della Fisica e Centro Studi e Ricerche ``Enrico Fermi', Rome, Italy
\item \Idef{org12}Chicago State University, Chicago, Illinois, United States
\item \Idef{org13}China Institute of Atomic Energy, Beijing, China
\item \Idef{org14}Chonbuk National University, Jeonju, Republic of Korea
\item \Idef{org15}Comenius University Bratislava, Faculty of Mathematics, Physics and Informatics, Bratislava, Slovakia
\item \Idef{org16}COMSATS Institute of Information Technology (CIIT), Islamabad, Pakistan
\item \Idef{org17}Creighton University, Omaha, Nebraska, United States
\item \Idef{org18}Department of Physics, Aligarh Muslim University, Aligarh, India
\item \Idef{org19}Department of Physics, Ohio State University, Columbus, Ohio, United States
\item \Idef{org20}Department of Physics, Pusan National University, Pusan, Republic of Korea
\item \Idef{org21}Department of Physics, Sejong University, Seoul, Republic of Korea
\item \Idef{org22}Department of Physics, University of California, Berkeley, California, United States
\item \Idef{org23}Department of Physics, University of Oslo, Oslo, Norway
\item \Idef{org24}Department of Physics and Technology, University of Bergen, Bergen, Norway
\item \Idef{org25}Dipartimento di Fisica dell'Universit\`{a} 'La Sapienza' and Sezione INFN, Rome, Italy
\item \Idef{org26}Dipartimento di Fisica dell'Universit\`infn{a} and Sezione INFN, Cagliari, Italy
\item \Idef{org27}Dipartimento di Fisica dell'Universit\`{a} and Sezione INFN, Trieste, Italy
\item \Idef{org28}Dipartimento di Fisica dell'Universit\`{a} and Sezione INFN, Turin, Italy
\item \Idef{org29}Dipartimento di Fisica e Astronomia dell'Universit\`{a} and Sezione INFN, Bologna, Italy
\item \Idef{org30}Dipartimento di Fisica e Astronomia dell'Universit\`{a} and Sezione INFN, Catania, Italy
\item \Idef{org31}Dipartimento di Fisica e Astronomia dell'Universit\`{a} and Sezione INFN, Padova, Italy
\item \Idef{org32}Dipartimento di Fisica `E.R.~Caianiello' dell'Universit\`{a} and Gruppo Collegato INFN, Salerno, Italy
\item \Idef{org33}Dipartimento DISAT del Politecnico and Sezione INFN, Turin, Italy
\item \Idef{org34}Dipartimento di Scienze e Innovazione Tecnologica dell'Universit\`{a} del Piemonte Orientale and INFN Sezione di Torino, Alessandria, Italy
\item \Idef{org35}Dipartimento Interateneo di Fisica `M.~Merlin' and Sezione INFN, Bari, Italy
\item \Idef{org36}European Organization for Nuclear Research (CERN), Geneva, Switzerland
\item \Idef{org37}Faculty of Engineering and Business Administration, Western Norway University of Applied Sciences, Bergen, Norway
\item \Idef{org38}Faculty of Nuclear Sciences and Physical Engineering, Czech Technical University in Prague, Prague, Czech Republic
\item \Idef{org39}Faculty of Science, P.J.~\v{S}af\'{a}rik University, Ko\v{s}ice, Slovakia
\item \Idef{org40}Frankfurt Institute for Advanced Studies, Johann Wolfgang Goethe-Universit\"{a}t Frankfurt, Frankfurt, Germany
\item \Idef{org41}Gangneung-Wonju National University, Gangneung, Republic of Korea
\item \Idef{org42}Gauhati University, Department of Physics, Guwahati, India
\item \Idef{org43}Helmholtz-Institut f\"{u}r Strahlen- und Kernphysik, Rheinische Friedrich-Wilhelms-Universit\"{a}t Bonn, Bonn, Germany
\item \Idef{org44}Helsinki Institute of Physics (HIP), Helsinki, Finland
\item \Idef{org45}Hiroshima University, Hiroshima, Japan
\item \Idef{org46}Hochschule Worms, Zentrum  f\"{u}r Technologietransfer und Telekommunikation (ZTT), Worms, Germany
\item \Idef{org47}Horia Hulubei National Institute of Physics and Nuclear Engineering, Bucharest, Romania
\item \Idef{org48}Indian Institute of Technology Bombay (IIT), Mumbai, India
\item \Idef{org49}Indian Institute of Technology Indore, Indore, India
\item \Idef{org50}Indonesian Institute of Sciences, Jakarta, Indonesia
\item \Idef{org51}INFN, Laboratori Nazionali di Frascati, Frascati, Italy
\item \Idef{org52}INFN, Sezione di Bari, Bari, Italy
\item \Idef{org53}INFN, Sezione di Bologna, Bologna, Italy
\item \Idef{org54}INFN, Sezione di Cagliari, Cagliari, Italy
\item \Idef{org55}INFN, Sezione di Catania, Catania, Italy
\item \Idef{org56}INFN, Sezione di Padova, Padova, Italy
\item \Idef{org57}INFN, Sezione di Roma, Rome, Italy
\item \Idef{org58}INFN, Sezione di Torino, Turin, Italy
\item \Idef{org59}INFN, Sezione di Trieste, Trieste, Italy
\item \Idef{org60}Inha University, Incheon, Republic of Korea
\item \Idef{org61}Institut de Physique Nucl\'{e}aire d'Orsay (IPNO), Institut National de Physique Nucl\'{e}aire et de Physique des Particules (IN2P3/CNRS), Universit\'{e} de Paris-Sud, Universit\'{e} Paris-Saclay, Orsay, France
\item \Idef{org62}Institute for Nuclear Research, Academy of Sciences, Moscow, Russia
\item \Idef{org63}Institute for Subatomic Physics of Utrecht University, Utrecht, Netherlands
\item \Idef{org64}Institute for Theoretical and Experimental Physics, Moscow, Russia
\item \Idef{org65}Institute of Experimental Physics, Slovak Academy of Sciences, Ko\v{s}ice, Slovakia
\item \Idef{org66}Institute of Physics, Bhubaneswar, India
\item \Idef{org67}Institute of Physics of the Czech Academy of Sciences, Prague, Czech Republic
\item \Idef{org68}Institute of Space Science (ISS), Bucharest, Romania
\item \Idef{org69}Institut f\"{u}r Kernphysik, Johann Wolfgang Goethe-Universit\"{a}t Frankfurt, Frankfurt, Germany
\item \Idef{org70}Instituto de Ciencias Nucleares, Universidad Nacional Aut\'{o}noma de M\'{e}xico, Mexico City, Mexico
\item \Idef{org71}Instituto de F\'{i}sica, Universidade Federal do Rio Grande do Sul (UFRGS), Porto Alegre, Brazil
\item \Idef{org72}Instituto de F\'{\i}sica, Universidad Nacional Aut\'{o}noma de M\'{e}xico, Mexico City, Mexico
\item \Idef{org73}iThemba LABS, National Research Foundation, Somerset West, South Africa
\item \Idef{org74}Johann-Wolfgang-Goethe Universit\"{a}t Frankfurt Institut f\"{u}r Informatik, Fachbereich Informatik und Mathematik, Frankfurt, Germany
\item \Idef{org75}Joint Institute for Nuclear Research (JINR), Dubna, Russia
\item \Idef{org76}Konkuk University, Seoul, Republic of Korea
\item \Idef{org77}Korea Institute of Science and Technology Information, Daejeon, Republic of Korea
\item \Idef{org78}KTO Karatay University, Konya, Turkey
\item \Idef{org79}Laboratoire de Physique Subatomique et de Cosmologie, Universit\'{e} Grenoble-Alpes, CNRS-IN2P3, Grenoble, France
\item \Idef{org80}Lawrence Berkeley National Laboratory, Berkeley, California, United States
\item \Idef{org81}Lund University Department of Physics, Division of Particle Physics, Lund, Sweden
\item \Idef{org82}Nagasaki Institute of Applied Science, Nagasaki, Japan
\item \Idef{org83}Nara Women{'}s University (NWU), Nara, Japan
\item \Idef{org84}National and Kapodistrian University of Athens, School of Science, Department of Physics , Athens, Greece
\item \Idef{org85}National Centre for Nuclear Research, Warsaw, Poland
\item \Idef{org86}National Institute of Science Education and Research, HBNI, Jatni, India
\item \Idef{org87}National Nuclear Research Center, Baku, Azerbaijan
\item \Idef{org88}National Research Centre Kurchatov Institute, Moscow, Russia
\item \Idef{org89}Niels Bohr Institute, University of Copenhagen, Copenhagen, Denmark
\item \Idef{org90}Nikhef, National institute for subatomic physics, Amsterdam, Netherlands
\item \Idef{org91}NRC ¿Kurchatov Institute¿ ¿ IHEP , Protvino, Russia
\item \Idef{org92}NRNU Moscow Engineering Physics Institute, Moscow, Russia
\item \Idef{org93}Nuclear Physics Group, STFC Daresbury Laboratory, Daresbury, United Kingdom
\item \Idef{org94}Nuclear Physics Institute of the Czech Academy of Sciences, \v{R}e\v{z} u Prahy, Czech Republic
\item \Idef{org95}Oak Ridge National Laboratory, Oak Ridge, Tennessee, United States
\item \Idef{org96}Petersburg Nuclear Physics Institute, Gatchina, Russia
\item \Idef{org97}Physics department, Faculty of science, University of Zagreb, Zagreb, Croatia
\item \Idef{org98}Physics Department, Panjab University, Chandigarh, India
\item \Idef{org99}Physics Department, University of Jammu, Jammu, India
\item \Idef{org100}Physics Department, University of Rajasthan, Jaipur, India
\item \Idef{org101}Physikalisches Institut, Eberhard-Karls-Universit\"{a}t T\"{u}bingen, T\"{u}bingen, Germany
\item \Idef{org102}Physikalisches Institut, Ruprecht-Karls-Universit\"{a}t Heidelberg, Heidelberg, Germany
\item \Idef{org103}Physik Department, Technische Universit\"{a}t M\"{u}nchen, Munich, Germany
\item \Idef{org104}Research Division and ExtreMe Matter Institute EMMI, GSI Helmholtzzentrum f\"ur Schwerionenforschung GmbH, Darmstadt, Germany
\item \Idef{org105}Rudjer Bo\v{s}kovi\'{c} Institute, Zagreb, Croatia
\item \Idef{org106}Russian Federal Nuclear Center (VNIIEF), Sarov, Russia
\item \Idef{org107}Saha Institute of Nuclear Physics, Kolkata, India
\item \Idef{org108}School of Physics and Astronomy, University of Birmingham, Birmingham, United Kingdom
\item \Idef{org109}Secci\'{o}n F\'{\i}sica, Departamento de Ciencias, Pontificia Universidad Cat\'{o}lica del Per\'{u}, Lima, Peru
\item \Idef{org110}Stefan Meyer Institut f\"{u}r Subatomare Physik (SMI), Vienna, Austria
\item \Idef{org111}SUBATECH, IMT Atlantique, Universit\'{e} de Nantes, CNRS-IN2P3, Nantes, France
\item \Idef{org112}Suranaree University of Technology, Nakhon Ratchasima, Thailand
\item \Idef{org113}Technical University of Ko\v{s}ice, Ko\v{s}ice, Slovakia
\item \Idef{org114}Technische Universit\"{a}t M\"{u}nchen, Excellence Cluster 'Universe', Munich, Germany
\item \Idef{org115}The Henryk Niewodniczanski Institute of Nuclear Physics, Polish Academy of Sciences, Cracow, Poland
\item \Idef{org116}The University of Texas at Austin, Austin, Texas, United States
\item \Idef{org117}Universidad Aut\'{o}noma de Sinaloa, Culiac\'{a}n, Mexico
\item \Idef{org118}Universidade de S\~{a}o Paulo (USP), S\~{a}o Paulo, Brazil
\item \Idef{org119}Universidade Estadual de Campinas (UNICAMP), Campinas, Brazil
\item \Idef{org120}Universidade Federal do ABC, Santo Andre, Brazil
\item \Idef{org121}University College of Southeast Norway, Tonsberg, Norway
\item \Idef{org122}University of Cape Town, Cape Town, South Africa
\item \Idef{org123}University of Houston, Houston, Texas, United States
\item \Idef{org124}University of Jyv\"{a}skyl\"{a}, Jyv\"{a}skyl\"{a}, Finland
\item \Idef{org125}University of Liverpool, Liverpool, United Kingdom
\item \Idef{org126}University of Split, Faculty of Electrical Engineering, Mechanical Engineering and Naval Architecture, Split, Croatia
\item \Idef{org127}University of Tennessee, Knoxville, Tennessee, United States
\item \Idef{org128}University of the Witwatersrand, Johannesburg, South Africa
\item \Idef{org129}University of Tokyo, Tokyo, Japan
\item \Idef{org130}University of Tsukuba, Tsukuba, Japan
\item \Idef{org131}Universit\'{e} Clermont Auvergne, CNRS/IN2P3, LPC, Clermont-Ferrand, France
\item \Idef{org132}Universit\'{e} de Lyon, Universit\'{e} Lyon 1, CNRS/IN2P3, IPN-Lyon, Villeurbanne, Lyon, France
\item \Idef{org133}Universit\'{e} de Strasbourg, CNRS, IPHC UMR 7178, F-67000 Strasbourg, France, Strasbourg, France
\item \Idef{org134} Universit\'{e} Paris-Saclay Centre d¿\'Etudes de Saclay (CEA), IRFU, Department de Physique Nucl\'{e}aire (DPhN), Saclay, France
\item \Idef{org135}Universit\`{a} degli Studi di Pavia, Pavia, Italy
\item \Idef{org136}Universit\`{a} di Brescia, Brescia, Italy
\item \Idef{org137}V.~Fock Institute for Physics, St. Petersburg State University, St. Petersburg, Russia
\item \Idef{org138}Variable Energy Cyclotron Centre, Kolkata, India
\item \Idef{org139}Warsaw University of Technology, Warsaw, Poland
\item \Idef{org140}Wayne State University, Detroit, Michigan, United States
\item \Idef{org141}Westf\"{a}lische Wilhelms-Universit\"{a}t M\"{u}nster, Institut f\"{u}r Kernphysik, M\"{u}nster, Germany
\item \Idef{org142}Wigner Research Centre for Physics, Hungarian Academy of Sciences, Budapest, Hungary
\item \Idef{org143}Yale University, New Haven, Connecticut, United States
\item \Idef{org144}Yonsei University, Seoul, Republic of Korea
\item \Idef{org145}Universita e INFN, Salerno, Italy
\end{Authlist}
\endgroup
%
\end{document}